\def\isarxivversion{1} 

\ifdefined\isarxivversion
\documentclass[11pt]{article}

\usepackage[numbers]{natbib}

\else

\documentclass[twoside]{article}
\usepackage{aistats2023}
\usepackage[round]{natbib}

\fi

\usepackage{amsmath}
\usepackage{amsthm}
\usepackage{amssymb}
\usepackage{algorithm}
\usepackage{subfig}
\usepackage{algpseudocode}
\usepackage{graphicx}
\usepackage{grffile}
\usepackage{wrapfig,epsfig}
\usepackage{url}
\usepackage{xcolor}
\usepackage{epstopdf}

\usepackage{bbm}
\usepackage{dsfont}

\allowdisplaybreaks

\ifdefined\isarxivversion

\let\C\relax
\usepackage{tikz}
\usepackage{hyperref}  
\hypersetup{colorlinks=true,citecolor=blue,linkcolor=blue} 
\usetikzlibrary{arrows}
\usepackage[margin=1in]{geometry}

\else

\usepackage{microtype}
\usepackage{hyperref}
\definecolor{mydarkblue}{rgb}{0,0.08,0.45}
\hypersetup{colorlinks=true, citecolor=mydarkblue,linkcolor=mydarkblue}

\fi

\newtheorem{theorem}{Theorem}[section]
\newtheorem{lemma}[theorem]{Lemma}
\newtheorem{definition}[theorem]{Definition}

\newtheorem{fact}[theorem]{Fact}
\newtheorem{remark}[theorem]{Remark}
\newtheorem{claim}[theorem]{Claim}

\newcommand{\wh}{\widehat}
\newcommand{\wt}{\widetilde}

\newcommand{\R}{\mathbb{R}}
\newcommand{\G}{\mathsf{G}}

\renewcommand{\tilde}{\wt}
\renewcommand{\hat}{\wh}

\DeclareMathOperator*{\E}{{\mathbb{E}}}

\DeclareMathOperator*{\Z}{\mathbb{Z}}
\DeclareMathOperator*{\C}{\mathbb{C}}

\DeclareMathOperator{\OPT}{OPT}

\DeclareMathOperator{\poly}{poly}

\DeclareMathOperator{\dist}{dist}
\DeclareMathOperator{\cost}{cost}

\DeclareMathOperator{\eps}{\epsilon} 

\makeatletter
\newcommand*{\RN}[1]{\expandafter\@slowromancap\romannumeral #1@}
\makeatother

\newcommand{\Yuanyuan}[1]{{\color{purple}[Yuanyuan: #1]}}

\usepackage{lineno}

\begin{document}

\ifdefined\isarxivversion

\date{}

\title{A Nearly Optimal Size Coreset Algorithm with Nearly Linear Time}
\author{
Yichuan Deng\thanks{\texttt{admindeng@mail.ustc.edu.cn}. USTC.} 
\and 
Zhao Song\thanks{\texttt{zsong@adobe.com}. Adobe Research.}
\and 
Yitan Wang\thanks{\texttt{yitan.wang@yale.edu}. Yale University. Yitan Wang gratefully acknowledges support from ONR Award N00014-20-1-2335.}
\and 
Yuanyuan Yang\thanks{\texttt{yyangh@cs.washington.edu}. University of Washington.}
}

\else

\twocolumn[

\aistatstitle{A Nearly Optimal Size Coreset Algorithm with Nearly Linear Time}

\aistatsauthor{ Yichuan Deng\thanks{\texttt{admindeng@mail.ustc.edu.cn}. USTC.} 
\and 
Zhao Song\thanks{\texttt{zsong@adobe.com}. Adobe Research.}
\and 
Yitan Wang\thanks{\texttt{yitan.wang@yale.edu}. Yale University.}
\and 
Yuanyuan Yang\thanks{\texttt{yyangh@cs.washington.edu}. University of Washington.} }

\aistatsaddress{ USTC \And  Adobe Research. \And Yale University \And University of Washington } ]

\fi

\ifdefined\isarxivversion
\begin{titlepage}
  \maketitle
  \begin{abstract}
A coreset is a point set containing information about geometric properties of a larger point set.
A series of previous works show that in many machine learning problems, especially in clustering problems, coreset could be very useful to build efficient algorithms.
Two main measures of an coreset construction algorithm's performance are the running time of the algorithm and the size of the coreset output by the algorithm.
In this paper we study the construction of coresets for the $(k,z)$-clustering problem, which is a generalization of $k$-means and $k$-median problem.
By properly designing a sketching-based distance estimation data structure, we propose faster algorithms that construct coresets with matching size of the state-of-the-art results.

  \end{abstract}
  \thispagestyle{empty}
\end{titlepage}

{\hypersetup{linkcolor=black}
\tableofcontents
}
\newpage

\else

\begin{abstract}
A coreset is a point set containing information about geometric properties of a larger point set.
A series of previous works show that in many machine learning problems, especially in clustering problems, coreset could be very useful to build efficient algorithms.
Two main measures of an coreset construction algorithm's performance are the running time of the algorithm and the size of the coreset output by the algorithm.
In this paper we study the construction of coresets for the $(k,z)$-clustering problem, which is a generalization of $k$-means and $k$-median problem.
By properly designing a sketching-based distance estimation data structure, we propose faster algorithms that construct coresets with matching size of the state-of-the-art results.
\end{abstract}

\fi

\section{Introduction}
With the ever-changing technology progress and rapid evolution of machine learning techniques, the magnitude of the data collected is growing greater during the recent years. 
Consequently, it is common in modern machine learning and data science applications that the size of the dataset used is extremely large. Since the running time and computational resources usually grows with the input size, efficient data preprocessing techniques are in great demand.
Therefore, a natural desire arises: we hope to enjoy the variety and the information brought by the large dataset, at the same time, we want to keep the volume of the data fed into algorithms small. Many machine learning algorithms work in the way of extracting information from the geometry of the data. Such observation motivates us to compress the size of the dataset with the constraint of preserving some geometric properties of the original dataset. If one could construct a smaller set of data points as an abstraction of the original dataset, i.e., approximate the same geometric properties of the large dataset, then implementing many machine learning algorithms over this abstraction would give similar results, while significantly reducing the overall consumed running time and spaces.

One seminal formalization of approximating data with geometric properties preserved is the concept of coreset~\citep{ahv+05}.
Intuitively, suppose $U\subseteq \R^d$ is a point set and we say a set $V \subseteq \R^d$ is a coreset of $U$ if queries on $V$ are approximately the same as queries on $U$ up to some precision parameter $\epsilon > 0$.
The different types of queries corresponds to the choice of the measure function $\mu$ over $\R^d$ in~\cite{ahv+05}. Since its formalization, the coreset has wide applications in numerical linear algebra~(\cite{cw09}), clustering~(\cite{ap00, hsv02, pjm+02,drvw06, fss13}), machine learning~(\cite{am04,blk15}), and subspace approximation~(\cite{drvw06, dv07, sv07,fmsw10, fl11, vx12, fss13}). 

As aforementioned, the coreset has been widely applied in machine learning algorithms.
One of well established applications of the coreset is clustering. The high level idea of clustering is to split data points into groups, with the hope of maintaining two properties: 1).\ the points in the \emph{same} cluster should behave similarly, and 2).\ the points in \emph{different} clusters are disparate.
Two crucial components of clustering algorithms are $k$-means and $k$-median.
The $k$-means method, $k$-median method and their variants have been studied for decades and used in many applications~(\cite{bhsi02, hm04, kmn+04, fs05, hk05, c06, fms07, hk07, fs08, c09, mnv09, fl11, fs12, fss13, ckm16, kls18, hsyz18,emz19,hv20,cmz22,emmz22}).
The discussion of previous works particularly related to this paper is referred to Section~\ref{sec:related_works}.

In this paper, we focus on a generalized clustering problem named $(k,z)$-clustering, which covers the $k$-means and $k$-median as special cases.
Suppose $U \subseteq \R^d$ is the set of $n$ collected data points in $d$-dimensional Euclidean space, $k \in \mathbb{Z}_{>0}$ is a positive integer, and $z \in [1,+\infty)$ is a fixed constant.
The $(k,z)$-clustering problem is to find a set $C \subseteq \R^d$ with size $k$, which is called the center set, such that minimizes 
\begin{align}\label{eqn:definition_of_cost}
    \cost_z(U, C) = \sum_{x \in U} d_z(x, C)
\end{align}
where $d_z(x,C):=\min_{c \in C} \|x-c\|^z_2$ denotes the $z$-th power of distance from $x$ to the center set $C$.
The $k$-means problem is equivalent to $(k,2)$-clustering problem and the $k$-median problem is equivalentt to $(k,1)$-clustering problem.

To efficiently solve the $(k,z)$-clustering problem, we follow the strategy of computing coreset for the dataset $U$. We adopt the standard definition of coreset in recent literatures~(\cite{hm04,hk07,fms07,ls10,fl11,blk17,hv20,bbh+20,css21}).
For a point set $U \subseteq \R^d$, we define the $\epsilon$-coreset as follows:
\begin{definition}[$\epsilon$-Coreset] 
\label{def:eps_coreset}
    Given a set $U \subseteq \R^d$ of $n$ points and $\epsilon \in (0, 0.1)$, we say set $D\subseteq \R^d$ is an $\epsilon$-coreset with weight function $w:D \rightarrow \R_+$  for $(k, z)$-clustering problem if it holds that for any  center set $C \subseteq \R^d$ of size $k$, the $(k, z)$-clustering cost with respected to $C$ is $\epsilon$-approximately preserved, i.e.,
    \begin{align*}
        1 - \epsilon  \le  \frac{\sum_{x \in D}w(x) \cdot d_z(x, C)}{ \cost_z(U, C)} \le  1 + \epsilon.
    \end{align*}
\end{definition}

Recall our motivation to find a coreset is to preserve the geometric properties of the original dataset and reduce the size of input for the following steps. Obviously, the preservation of the geometric properties is characterized by the definition of the coreset itself. Therefore, it is natural to measure the performance of an algorithm on constructing the coreset in two aspects: 1).\ the size of the coreset output by the constructing algorithm, and 2).\ the running time of the constructing algorithm. In this light, the goal of the coreset construction is to design a \emph{faster} algorithm that outputs a \emph{smaller} coreset, with the same approximation guarantee. 
There have been many literature working on either optimizing the size of the coreset for certain precision parameter, or the time complexity for finding a coreset of a certain size.
Without any additional assumptions, it's often the case that the size of the coreset would have a dependency on the point dimension $d$.
However, for $k$-means and $k$-median problems, such dependency can be eliminated~\citep{fss13, bfl+16, sw18, bbcg19}. 

To the best of our knowledge, the state-of-the-art algorithm that solves $\epsilon$-coreset for $(k,z)$-clustering problem is proposed in~\cite{clss22}.
Via chaining techniques used by Gaussian processes, the algorithm developed in \cite{clss22} could compute a $\epsilon$-coreset with size $\wt{O}(\epsilon^{-2} \cdot k \cdot 2^{O(z \log z)} \cdot \min\{\epsilon^{-z}, k\})$ in $\wt{O}(ndk)$ time. 
Thus, it is natural ask the following question:
\begin{center}
    {\it Is it possible to compute an $\epsilon$-coreset for $(k,z)$-clustering problem in $o(ndk)$ time?}
\end{center}
In this work, we answer the above question affirmatively, by improving the runtime to $\tilde{O}(nd+nk)$.

\subsection{Related Works}\label{sec:related_works}
Many previous coreset algorithms and our coreset algorithm take center set generation as a subroutine. \cite{mp04} introduce an efficient algorithm named successive sampling for the center set generation in $k$-median setting.
The successive sampling algorithm takes a set containing $n$ points in $\R^d$ as input and with high probability constructs an assignment function $\sigma: U \rightarrow U$ such that $|\sigma(U)| \le k$ and $\sum_{x\in U} d_1(x,\sigma(x)) w(x) = O(\mathrm{OPT}_k)$ where $\mathrm{OPT}_k$ is defined as $\mathrm{OPT}_k = \min_{C:|C|\le k} \cost_1(U,C)$.
The authors prove that the successive sampling algorithm runs in $\wt{O}(ndk)$ time.

\cite{sw18} come up an algorithm constructing coresets for the k-median and subspace approximation.
The size of $\epsilon$-coreset constructed by their method is $\poly(k/\epsilon)$, which is independent of the input set's size $n$ and the dimension of the Euclidean space $d$. The running time of their algorithm is $\wt{O}(nd+2^{\poly(k)})$.

\cite{hv20} develop an importance sampling algorithm to construct $\epsilon$-coresets for $(k,z)$-clustering problem.
By the two-stage importance sampling technique developed in their paper, the algorithm outputs an $\epsilon$-coreset of size $\wt{O}(\epsilon^{-\Omega(z)}k)$ in time $\wt{O}(ndk)$.

\cite{css21} give an method to construct an $\epsilon$-coreset for $(k,z)$-clustering problem in time $\wt{O}(ndk)$ and the size of the coreset is
$O(\frac{2^{O(z \log z) } \cdot \log^4(1/\epsilon)}{\min\{\epsilon^2, \epsilon^z\}}\cdot (k \log k + \log\log(1/\epsilon)+\log(1 / \delta)))$.
We summarize these related results in Table~\ref{tab:history_of_results}.

\begin{table*}[!ht]
    \centering
    \begin{tabular}{|l|l|l|l|} \hline 
        {\bf References} & {\bf Objective} & {\bf Coreset Size} & {\bf Time}  \\ \hline 
         \cite{fl11} & $(k,z)$-cluster & $\epsilon^{-2z}k d \log(k/\epsilon)$ & $ndk$ \\ \hline 
         \cite{vx12} & $(k,z)$-cluster & $2^{2z} \epsilon^{-2} k d \log(k/\epsilon)$ & $\poly(n,d)$ \\ \hline 
         \cite{sw18} & $(k,z)$-cluster & $\poly(k/\epsilon^z)$ & $nd+2^{\poly(k)}$\\ \hline 
         \cite{hv20}* & $(k,z)$-cluster & $\min\{\epsilon^{-2z-2}, 2^{2z}\epsilon^{-4}k\} \cdot k \log k \log(k/\epsilon)$ & $ndk$ \\ \hline 
         \cite{css21} & $(k, z)$-cluster  & $\frac{2^{O(z \log z) } \cdot \log^4(1/\epsilon)}{\min\{\epsilon^2, \epsilon^z\}}\cdot (k \log k + \log\log(1/\epsilon)+\log(1 / \delta))$ & $n d k$ \\ \hline
         \cite{bvj+21} & $(k, z)$-cluster  & $\epsilon^{-4} \cdot k^2 \cdot 2^{2z} \cdot \log(1/\delta)$ & $n d k$ \\ \hline
         \cite{clss22} & $(k, z)$-cluster  & $\epsilon^{-2}\cdot k \cdot 2^{O(z \log z)} \cdot \min\{\epsilon^{-z}, k\}$ & $n d k$ \\ \hline
         Ours &  $(k,z)$-cluster & $\epsilon^{-2z-2} \cdot k \log k \log(k/\epsilon)$ & $nd+nk$ \\ \hline \hline 
         \cite{fl11} & $k$-median & $\epsilon^{-2} \cdot k d \log k$ & $ndk$
         \\ \hline 
         \cite{sw18} & $k$-median & $\epsilon^{-4} \cdot k^2  \log k$ & $nd + 2^{\poly(k)}$ \\ \hline 
         \cite{hv20}* & $k$-median & $\epsilon^{-4} \cdot k \log k \log(k/\epsilon)$ & $ndk$ \\ \hline 
         \cite{css21} & $k$-median  & $\epsilon^{-2} \cdot \log^4(1/\epsilon)\cdot (k \log k + \log\log(1/\epsilon)+\log(1 / \delta))$ & $n d k$ \\ \hline
         \cite{bvj+21} & $k$-median  & $\epsilon^{-4} \cdot k^2 \cdot \log(1/\delta)$ & $n d k$ \\ \hline
        \cite{clss22} & $k$-median  & $\epsilon^{-2}\cdot k \cdot \min\{\epsilon^{-1}, k\}$ & $n d k$ \\ \hline
         Ours & $k$-median & $\epsilon^{-4} \cdot k \log k \log(k/\epsilon)$  & $nd+nk$ \\ \hline \hline
         \cite{bfl+16} & $k$-means & $\epsilon^{-3} \cdot k^2\log(k/\epsilon)$ & $n d k$\\ \hline 
         \cite{bbcg19} & $k$-means & $\epsilon^{-8} \cdot k\log^2(k/\epsilon)\log(1/\epsilon)$ & $\poly(n,d)$ \\ \hline 
         \cite{hv20}* & $k$-means & $\epsilon^{-6}\cdot k\log k \log(k/\epsilon)$ & $ndk$ \\ \hline 
         \cite{css21} & $k$-means  & $\epsilon^{-2} \cdot \log^4(1/\epsilon)\cdot (k \log k + \log\log(1/\epsilon)+\log(1 / \delta))$ & $n d k$
         \\ \hline
         \cite{bvj+21} & $k$-means  & $\epsilon^{-4} \cdot k^2 \cdot \log(1/\delta)$ & $n d k$ \\ \hline
         \cite{clss22} & $k$-means  & $\epsilon^{-2}\cdot k \cdot \min\{\epsilon^{-2}, k\}$ & $n d k$ \\ \hline
         Ours & $k$-means & $\epsilon^{-6}\cdot k\log k \log(k/\epsilon)$ & $nd+nk$ \\ \hline 
    \end{tabular}
    \caption{The $n$ denotes the size of input set $U$. The $(k,z)$ denote the parameters for clustering problem. The $\delta$ denotes the failure probability for randomized algorithms. The $\epsilon$ denote the approximation error. For simplicity, we ignore the big-Oh in the table. The objective denote the optimization for $(k,z)$-clustering, k-median and k-means, respectively. We treat $z$ as constant in running time. For the ``Time'' column, we ignore the poly log factors for simplicity. The references marked with * have the same coreset size as ours. Almost all the existing running time for $(k,z)$-clustering, k-median and k-means are in $O(ndk$), except for ~\cite{sw18}.
    The size of the coreset output by our algorithm matches the result in \cite{hv20} and our algorithm is the first one running in $\wt{O}(nd+nk)$ time. }
    \label{tab:history_of_results}
\end{table*}

\subsection{Our Contribution}
In this work, our contributions consist of two parts.
First, we develop two randomized algorithms that outputs a coreset with matching size of previous works in smaller running time. Table~\ref{tab:history_of_results} shows the comparison between our results and previous works. Second, one by-product of our work is a faster algorithm computing center set.

Our first contribution is an algorithm generating an $\epsilon$-corset of size $\epsilon^{ -\Theta(z)} \cdot \wt{O}(k)$ in time $\wt{O}( nd + nk )$.
\begin{theorem}[Informal Version of Combination of Theorem~\ref{thm:main_correct} and Theorem~\ref{thm:main_runtime}]
    Given a dataset $U$ of $n$ points in $\R^d$, a coreset parameter $\epsilon \in (0, 0.1)$, 
    and a failure probability $\delta \in (0, 0.1)$, for any constant $z \geq 1$, there is an algorithm (Algorithm~\ref{alg:coreset_gen_alg}) that outputs an $\epsilon$-coreset for the $(k, z)$-clustering problem of size  
    \begin{align*}
        \epsilon^{ -\Theta(z)} \cdot \wt{O}(k) ,
    \end{align*}
    with probability at least $1 - \delta$, and runs in time
    \begin{align*}
        \wt{O}( nd + nk ).
    \end{align*}
\end{theorem}
We claim our running time $\wt{O}(nd+nk)$ is nearly optimal up to logarithmic factors in the sense that $\Omega(nd)$ is necessary for computing an $\epsilon$-coreset of a set containing $n$ points in $d$-dimensional Euclidean space.
Reading all points in the input set $U$ takes $\Omega(nd)$ time and thus is a trivial lower bound for the running time.
In high dimensional setting, typical cases are that the number of sample points is much less than the dimension, i.e., $n \ll d$.
It is also clear that for $(k,z)$-clustering problem, we only focus on the cases where $k < n$ because when $k \ge n$ the trivial optimal solution $C^*$ minimizing $\cost_z(U,C)$ is $C^*=U$.
Hence in the high dimensional case, $\wt{O}(nd+nk)=\wt{O}(nd)$ and the running time of our algorithm is optimal up to logarithmic factors.
Even if we do not assume the high dimensional condition, we conjecture that $\Omega(nk)$ is still necessary.
The intuition of our conjecture is that there are $nk$ pairs of points $(x,c) \in U \times C$.
Even we could compute $\|x-c\|_2$ in $O(1)$ time, we conjecture the computation of $\cost_z(U,C)$ takes at least $\Omega(nk)$ time.

Our second contribution is an algorithm with running time $\wt{O}(nd+nk)$ that outputs an $\epsilon$-coreset of size matching the algorithm proposed in~\cite{css21}.
\begin{theorem}[Informal Version of Theorem~\ref{thm:correct_css21} and Theorem~\ref{thm:runtime_css21}]
    Let $X \subseteq \R^d$ be a dataset of $n$ points. Let $\epsilon \in (0, 0.1)$ be the coreset parameter. Let $\delta$ be the failure probability. There exists an algorithm (Algorithm~\ref{alg:coreset_gen_css21}) that outputs an $\epsilon$-coreset for the $(k,z)$-clustering problem of size  
    \begin{align*}
        O(\frac{2^{O(z \log z) } \cdot \log^4(1/\epsilon)}{\min\{\epsilon^2, \epsilon^z\}} \cdot (k \log k + \log\log(1/\epsilon)+\log(1 / \delta))),
    \end{align*}
    with probability at least $1 - \delta$, and runs in time
    \begin{align*}
        \wt{O}(n d + n k).
    \end{align*}
\end{theorem}

Our final result is an algorithm that computes a $(k,\Omega(1))$-center set (Definition~\ref{def:alhpa_center_set}) of the input in $\wt{O}(nd+nk)$ time.
\begin{theorem}[Informal version of Theorem~\ref{thm:center_set_gen_correct} and Theorem~\ref{thm:center_set_gen_runtime}]
    Given a set $U \subseteq \R^d$ of size $n$. Let $k \in \Z$ be the clustering parameter. There exists an algorithm that, it takes $U$ as an input, and outputs a $(k, \Omega(1))$-center set of $U$(Definition~\ref{def:alhpa_center_set}), and runs in time
    \begin{align*}
        \wt{O}(n k + n d).
    \end{align*}
\end{theorem}
\begin{remark}
    Our algorithm for center set generation is a generalization of that in \cite{mp04}. The running time is a direct improvement of \cite{mp04}, which runs in time $\wt{O}(n d k)$. Additionally, the generated center set is a subroutine of the final coreset generation. Hence, optimizing the running time for this subroutine helps reducing the overall running time for our algorithm. 
\end{remark} 

\section{Distance Estimation Data Structure}\label{sec:distance_estimation_data_structure}
In this section, we introduce the sketching-based distance estimation data structure, which is the main technique building block for our algorithms to achieve faster running time.
We formalize distance estimation problem (DEP) in section~\ref{sec:formulation_dep} and show the sketching-based data structure in section~\ref{sec:sketching_ds_dep}.

\subsection{Formulation of Distance Estimation Problem}\label{sec:formulation_dep}
A series of recent works studies the distance estimation problem~\citep{cn20,cn22,dswz22}.
The distance estimation problem asks us to maintain a set $X$ containing $n$ points in $d$-dimensional Euclidean space $C=\{c_1,\cdots,c_n\} \subseteq \R^d$.
In each time, we receive one operation in three types:
\begin{itemize}
    \item \textsc{Query}$(q)$ where $q \in \R^d$,
    \item \textsc{QueryMin}$(q,z)$ where $q\in\R^d$ and $z\in \R$,
    \item \textsc{Update}$(i,c)$ where $i\in[n]$ and $c \in \R^d$.
\end{itemize}
When we receive a \textsc{Query}$(q)$ operation, we need to return a list of distance $\{d_i\}_{i=1}^n$ where $d_i := \|c_i - q\|_2$ is the Euclidean distance between $c_i$ and $q$.
When the operation received is \textsc{QueryMin}$(q,z)$, we need to return the $z$-th power of the minimum distance between $q$ and $C$, i.e., return the value of $\min_{c\in C} \|c-q\|_2^z$.
Finally, if we receive an \textsc{Update}$(i,c)$ operation, then we change the $i$-th element in $C$ to $c$, i.e., $c_i \leftarrow c$, and make sure that $c_i$ is treated as the newly assigned value $c$ in the later operations until $c_i$ is modified again.

\subsection{Sketching-based Data Structure for Distance Estimation}\label{sec:sketching_ds_dep}
Obviously, a naive solution for the distance estimation problem is storing all points $\{c_i\}_{i=1}^n$ in an array.
For the \textsc{Update}$(i,c)$ operation, we simply assign the $i$-th element of the point array by $c$.
Since $c_i$ and $c$ are in $d$-dimensional Euclidean space and we represent them by a tuple of $d$ real numbers, one \textsc{Update}$(i,c)$ operation takes $O(d)$ time.
For the \textsc{Query}$(q)$ and \textsc{QueryMin}$(q,z)$ operations, the corresponding naive method is to enumerate all the pairs of $(c_i, q)$ and compute $\|c_i-q\|_2$, which takes $O(nd)$ time.

In this paper, instead of the naive implementation, we use a sketching-based data structure to accelerate the distance estimation problem via randomness.
The main idea of the sketching-based data structure used in this paper is inspired by the Johnson-Lindenstrauss transform~\citep{jl84}.
We first generate a random sketching matrix $\Pi \in \R^{m\times d}$, where $m$ is a parameter that we will discuss how to choose later.
By properly sampling the random matrix $\Pi$, we could make sure the distance between two points $\|x-y\|_2$ is well approximated by the distance between the sketch points $\|\Pi x - \Pi y\|_2$ even if we select $m$ far smaller than $d$.
Specifically, if we let $\Pi\in \R^{m\times d}$ be a random Gaussian matrix and let $m=\epsilon^{-2}\log(n/\delta)$ for some precision parameter $\epsilon > 0$ and some failure probability parameter, then for any fixed set of points $S$ with $S\subseteq \R^d$ and $|S|=n$, with probability at least $1-\delta$, it holds that for all points $x$ and $y$ in $S$,
\begin{align*}
    1-\epsilon \le \frac{\|\Pi x - \Pi y\|_2}{\|x-y\|_2} \le 1+\epsilon.
\end{align*}

Given that the random Gaussian matrix $\Pi$ preserves the distance well, we solve the distance estimation problem by storing an array of sketch points $\{\Pi c_i\}_{i=1}^n$.
When we receive \textsc{Query}$(q)$ and \textsc{QueryMin}$(q,z)$, we enumerate all pairs of $(\Pi c_i, q)$ and compute $\|\Pi c_i - \Pi q\|_2$.
As $m$ is smaller than $d$, computing $\|\Pi c_i - \Pi q\|_2$ can be done in less time than $\|c_i - q\|_2$.
Correspondingly, when we receive \textsc{Update}$(i,c)$, we store the sketch point $\Pi c$.

We formally state our results in the following two lemmas.
\begin{lemma}[Distance Estimation]
\label{lem:ds_dst_est_correct:main_text}
    Given a set of $n$ points in $\R^d$, an accuracy parameter $\epsilon \in (0, 0.1)$, and a failure probability $\delta \in (0, 0.1)$, there is a data structure (Algorithm~\ref{alg:data_structure_dist_est}) for the distance estimation problem such that the \textsc{Query} outputs $\{d_i\}_{i \in [n]}$ satisfing
    \begin{align*}
        \Pr\bigg[\forall i \in [n], 1 - \eps  \le \frac{d_i}{\|q - c_i\|_2} \le 1 + \eps\bigg] \ge 1 - \delta.
    \end{align*}
\end{lemma}
\begin{proof}
Due to the space limitation, we put details of Algorithm~\ref{alg:data_structure_dist_est} and the proof in the appendix section~\ref{sec:dist_est_ds}.
\end{proof}

\begin{lemma}[Running Time] 
\label{lem:ds_dst_est_runtime:main_text}
    Given a set of $n$ points in $\R^d$, for any accuracy parameter $\epsilon \in (0, 0.1)$
    , and any failure probability $\delta \in (0, 0.1)$, there is a data structure (Algorithm~\ref{alg:data_structure_dist_est}) using $O(\epsilon^{-2} (n + d) \log(n/\delta))$ spaces that supports the following operations:
    \begin{itemize}
        \item \textsc{Init}$(n, d, \delta, \epsilon, \{c_i\}_{i \in [n]})$ where $n \in \Z_+$, $d \in \Z_+$, $\delta \in (0, 0.1)$, $\epsilon \in (0,0.1)$, and $\{c_i\}_{i \in [n]} \subset \R^d$. It takes number of points $n$, dimension $d$, failure probability $\delta$, accuracy parameter $\eps$ and a set of points $\{c_i\}_{i \in [n]}$ as inputs. It runs in time 
        \begin{align*} 
            O(\epsilon^{-2} n d \log (n / \delta)).
        \end{align*}
        \item \textsc{Update}$(i \in [n], c \in \R^d)$. It takes index of point to be update $i$, and an update point $c$ as inputs. 
        It runs in time of $O(\epsilon^{-2} d \log (n / \delta))$,
        \item \textsc{Query}$(q \in \R^d)$. It takes a query point $q$ as input. 
        It runs in time of $O(\epsilon^{-2}(n + d)\log(n / \delta))$.
        \item \textsc{QueryMin}$(q \in \R^d, z \in \R)$. It takes a query point $q$ and a parameter $z$ as power of the distance as input. It runs in time of $O(\epsilon^{-2}(n + d)\log(n / \delta))$.
    \end{itemize}
\end{lemma}
\begin{proof}
See the appendix section~\ref{sec:dist_est_ds}.
\end{proof}

\section{Center Set Construction}\label{sec:center_set_construction}

In this section, we show how the sketching-based distance estimation data structure introduced in section~\ref{sec:distance_estimation_data_structure} helps us generate a center set in less running time.
The reason for making the effort to develop a faster center set constructing algorithm is that our coreset generating algorithm would take the center set construction as a subroutine.
Hence the coreset generation would benefit from a faster algorithm for center set construction.

\begin{algorithm*}[!ht]\caption{Center Set Generation, Informal version of Algorithm~\ref{alg:center_set_gen}}
\label{alg:center_set_gen:main_text}
    \begin{algorithmic}[1]
    \Procedure{CenterSetGen}{$\delta_0, n, d, U \subset \R^d, k, z, \eps_0$}
    \Comment{$|U|=n$}
        \State $i\gets 0$, $\quad U_0 \gets U$, $\quad V \gets \emptyset$, $\quad \delta_1 \gets O(\delta_0)$ \Comment{Initialize candidates and center set}
    	\State \textsc{DistanceEst} $D$
     	\Comment{data structure in Algo~\ref{alg:data_structure_dist_est}}
    	\State $\alpha \gets O(1)$, $\beta \gets O(1)$
    	\While{$|U_i| > \alpha k$} \label{line:center_gen_while_loop} \Comment{When there are enough ($O(k)$) candidates execute the loop}
    	    \State Sample a set $S_i$ from $U_i$ $\lfloor \alpha k \rfloor$ times with replacement, where for each time sample the points with equal probability. \label{line:center_gen_sampling}
    	   \State $D.\textsc{Init}(|S_i|, d, \delta_1, \epsilon_0/z, S_i)$ \label{line:center_gen_dist_init}
    	   \State $\wt{d}_x \gets D.\textsc{Query}(x)^z$ for all $x \in U_i$
    	    \State Compute $v_i$ using $\wt{d}_x$ where $v_i$ is the smallest radius for balls such that union of balls centered at points in $S_i$ covers at least $\beta|U_i|$ points in $U_i$. 
    	    \State $C_i \gets B(S_i, v_i) \cap U_i$ \Comment{$B(S,r)$ is the union of balls centered at points in $S$ with radius $r$.} 
    	    \State $U_{i+1} \gets U_i \backslash C_i$, $\quad V \gets V \cup S_i$, $\quad i\gets i+1$ \Comment{Update candidate and center set}
    	\EndWhile
    	\State $V \gets V \cup U_i$     \Comment{Include the last $O(k)$ candiates}
    	\State \Return $V$
    \EndProcedure
    \end{algorithmic}
\end{algorithm*}

\subsection{Formulation of Center Set Construction}
Let the cost between two point sets $U$ and $C$ be defined as Eq. \eqref{eqn:definition_of_cost}, i.e., $\cost_z(U,C)=\sum_{x\in U}d_z(x,C)$.
We say a set $V^*$ is a $(k,\alpha)$-center set for $U$ if $\cost_z(U,V^*)$ is approximately (up to constant $\alpha$) a lower bound for the value of $\inf\{\cost_z(U,C): V\subset \R^d, |C|=k\}$.
We formally define the $(k,\alpha)$-center set for $U$ as follows.
\begin{definition}[$(k, \alpha)$-Center Set]
\label{def:alhpa_center_set:main_text}
    Given $\alpha \in [1, +\infty)$, and a set $U \subset \R^d$ of $n$ points, we call a set $V^* \subset \R^d$ a $(k, \alpha)$-center set  
    for a given $U \subset \R^d$
    if it satisfies
    \begin{align*}
        \cost_z(U, V^*) \le \alpha \cdot \cost_z(U, C)
    \end{align*}
    for any other set $C$ with $k$ points.
\end{definition}
The goal of the center set construction problem is to find a $(k,\alpha)$-center set for $U$ where $U$, $k$, and $\alpha$ are given as input.
We emphasize that in the definition of $(k,\alpha)$-center set, the size of $V^*$ is \emph{not} required to be $k$.
In contrast, recall that the $(k,z)$-clustering problem requires the size of $C$ in equation \eqref{eqn:definition_of_cost} to be $k$.
Therefore, the $(k,z)$-clustering problem is \emph{not} equivalent to $(k,1)$-center set construction problem.
Similar to the performance measure of coreset generation algorithms, an algorithm which outputs center set with smaller size and runs in less time would be considered as a better one.

\subsection{Faster Center Set Construction}
As aforementioned, the center set construction is used as a subroutine for coreset generating in this paper.
Based on analysis we made for our coreset generation algorithm, it is sufficient to get a $(k, O(1))$-center set.
Thus in this section, we show a randomized algorithm constructing $(k,O(1))$-center set.

The intuition of the center set construction starts from a simple observation: $V^*=U$ is always a $(k,O(1))$-center set for $U$ since $\cost_z(U,U)=0$, which implies the inequality in definition~\ref{def:alhpa_center_set:main_text} always holds.
However, such choice of $V^*$ contains redundancy in the sense that it is not necessary to simultaneously include points closed to each other in $V^*$.
Suppose $p$ and $q$ are two points in $U$ and they are closed to each other, i.e., $\|p-q\|_2 \le \epsilon$ for some relatively small $\epsilon$.
Then by the triangle inequality, we know for all $x \in \R^d$, it holds that \begin{align*}
    \|x-p\|_2 - \epsilon \le \|x-q\|_2 \le \|x-p\|_2 + \epsilon.
\end{align*}
Hence $\|x-p\|_2$ is approximately the same as $\|x-q\|_2$.
If $p$ and $q$ are both contained in $V^*$, removing either $p$ or $q$ from $V^*$ would not hugely change the cost between $U$ and $V^*$.
Therefore by removing such redundant points in $V^*$ we could make the size of $V^*$ smaller.
One may note that the triangle inequality in the discussion above is for $\ell_2$-norm, which corresponds to $z=1$ for $d_z(x,y)=\|x-y\|_2$, and may thus worry about whether such intuition works for general $z$.
Fortunately, we have relaxed triangle inequality for general $z$ (Lemma~\ref{lem:relax_tri_ineq}) and the high level idea still works well for general $z$. 

With the intuition introduced above, we now describe the construction of $(k,O(1))$-center set.
The construction is an iterative procedure.
We maintain a candidate set of points and add into the center set.
Let $U_i$ denote the candidate set in the $i$-th iteration.
We begin with $U_0 = U$ and $V=\emptyset$ as every point in $U$ might be added into the center set and no point has been added into the center set.
In the $i$-th iteration, we sample a uniformly random subset $S_i$ from the candidate set $U_i$.
Such selected points in $S_i$ is then added into the center set $V$, i.e., $V\leftarrow V\cup S_i$.
Let $C_i$ denote the set of candidate points in $U_i$ which are closed to points in $S_i$.
We point out that since $C_i$ is defined as the set of points closed to the set $S_i$, the data structure for distance estimation problem introduced in section~\ref{sec:distance_estimation_data_structure} can be used in the computation of $C_i$ to reduce the time complexity.
According to the intuition of reducing redundancy, we would not add points in $C_i$ into the center set $V$ in any future iterations, since we have just added points in $S_i$ into the center set $V$.
Thus we eliminate points in $C_i$ from the candidate sets for next round, i.e., we assign $U_{i+1} \leftarrow U_i \backslash C_i$.
Due to the space limitation, we show an informal version (Algo~\ref{alg:center_set_gen:main_text}) of the center set construction algorithm. The formal version with more details are presented in appendix Algorithm~\ref{alg:center_set_gen}.

We present following two theorems as the analysis for correctness and running time of our center set algorithm.

\begin{theorem}[Center Set Generation Correctness]
\label{thm:center_set_gen_correct:main_text}
    Given a set $U$ of size $n$. 
    Let $c_0 > 1$ be a constant. Let $\delta_1 \in (0, 1)$ be a failure probability. The procedure \textsc{CenterSetGen} in Algorithm~\ref{alg:center_set_gen} output an $(k, O(1))$-center set $V$
    \begin{itemize}
        \item $|V| = O(k \log(n/k))$;
        \item $O(1) \cdot \cost_z(U,C) \geq \cost_z(U,V)$, for any $C \in \mathcal{C}_k$
    \end{itemize}
    with failure probability at most $\delta_0 = e^{-c_0 k} + \delta_1$, where $\mathcal{C}_k$ is defined as the domain of point sets of size $k$ in Euclidean space
    \begin{align*}
        \mathcal{C}_k := \{C = (c_1, \dots, c_k) ~|~ c_i \in \R^d, \forall i \in [k]\}.
    \end{align*}
\end{theorem}
\begin{proof}
See appendix section~\ref{sec:center_set_gen_correct}.
\end{proof}
\begin{theorem}[Center Set Generation Time]
\label{thm:center_set_gen_runtime}
    Given an $n$-point set in $\R^d$, an accuracy parameter $\epsilon_0 \in (0, 0. 1)$, and a failure probability $\delta_0 \in (0, 0.1)$, the Procedure \textsc{CenterSetGen} (Algorithm~\ref{alg:center_set_gen}) runs in time 
    \begin{align*}
        O( \epsilon_0^{-2} z^2 (n k + n d) \log(k/\delta_0) \log(n/k) ).
    \end{align*}
\end{theorem}
\begin{proof}
See appendix section~\ref{sec:center_set_gen_alg}.
\end{proof}
\begin{remark}
\cite{mp04} gives an algorithm constructing the center set in time $O(ndk\log(n/k)) = \wt{O}(ndk)$.
Our improvement for the running time of center set construction is necessary as the ultimate goal in this paper is to build faster coreset generating algorithm.
If we simply take the algorithm in \cite{mp04} as a subroutine for coreset generation then we can not achieve $\wt{O}(nd+nk)$ running time for the coreset algorithm.
We point out that many coreset algorithms take the center set algorithm as a subroutine, e.g., \cite{bvj+21} and \cite{clss22}.
So potentially these coreset algorithms involving center set construction could benefit from our faster center set generation.
\end{remark}

\section{Coreset Generation}

\begin{algorithm*}[!ht]\caption{Our algorithm for Coreset generating task}
\label{alg:coreset_gen_alg}
    \begin{algorithmic}[1]
    \Procedure{CoresetGen}{$U, n, d, \epsilon, \delta \in (0, 0.1), z \ge 1, k \ge 1$} \Comment{Theorem~\ref{thm:main_correct}, and ~\ref{thm:main_runtime}}
        \State $\gamma \gets O(1)$
        \State $C^* \gets \textsc{CenterSetGen}(\gamma,\delta, n, d, U, k, z)$ 
        \label{line:coreset_center_set_init} \label{line:center_set_gen}\Comment{Algorithm~\ref{alg:center_set_gen}}
        \State \textsc{DistanceEst} $\textsc{de}$
        \State $\textsc{de}.\textsc{Init}(k, d, \delta_1, \epsilon_1/z, C^*)$ \label{line:coreset_dist_init} \label{line:coreset_center_gen} \Comment{Algorithm~\ref{alg:data_structure_dist_est}} 
        \State $N \gets z^{O(z)} \cdot \epsilon^{-\Omega(z)} \cdot k^5 \cdot \log(k/\delta)$
        
        \For{$x \in U$} \label{line:query_loop_sketch_points} 
            \State $t \gets \textsc{de}.\textsc{QueryMin}(x,z)$ \label{line:coreset_find_min}
            \State $\wt{c}^*(x) \gets v_t$ \Comment{$\wt{c}^*(x)$ sends $x$ to its approx closest point in $C^*$}
        \EndFor
        
        \For{$v \in C^*$} 
            \State $X_{v} \gets \{x \in U ~|~  \wt{c}^*(x) = v\}$ \label{line:coreset_loop_set_gen}
        \EndFor
        
        \For{$x \in U$} 
    		    \State $\wt{\sigma}(x) \gets \frac{ \wt{d}_z(x, \wt{c}^*(x))}{ \wt{\cost}_z(U,C^*)} + \frac{1}{|X_{\wt{c}^*(x)}|}$\label{line:coreset_sigma_compute}
    		    \Comment{Definition~\ref{def:sigma_func_approx}}
    		\EndFor
    		\State For each $x \in U$, compute $p_x = \frac{\wt{\sigma}(x)}{\sum_{y \in U} \wt{\sigma}(y)}$
    		\State Let $D$ ($|D| = N$) be a subset sampled from $U$, where every $x \in U$ is sampled with prob. $p_x$, without replacement.  \label{line:coreset_sampling}
    		\For{$x \in D$}
    		    \State $w(x) \gets \frac{\sum_{y \in U} \wt{\sigma} (y)}{|D|\cdot \wt{\sigma}(x)}$ \label{line:coreset_compute_u} 
    		\EndFor
    		\State \Return $(D, w)$
    \EndProcedure
    \end{algorithmic}
\end{algorithm*}

In this section, we show our algorithm for coreset generation, which takes the distance estimation data structure presented in section~\ref{sec:distance_estimation_data_structure} and the center set construction algorithm introduced in section~\ref{sec:center_set_construction} as subroutines.

There are a series of previous works giving sampling-based algorithms to generate coreset.
The high-level procedure of the sampling-based algorithms are finding some weighting scores for each points in the input set $U$ and then generate the coreset by sampling points from $U$ according to the weighting scores.
For example, authors of \cite{fl11} use the sensitivity
\begin{align*}
    \sigma(x):=\sup_{C\subseteq U, |C|=k} \frac{d_z(x,C)}{\cost_z(U,C)}
\end{align*}
as the weighting score and authors of \cite{hv20} use a two-stage importance score sampling.
The key component of these sampling-based coreset generating algorithms is designing the weighting score function for points in $\R^d$.
The weighting score of $x$ should roughly characterize the difference of total cost between adding $x$ into a coreset and excluding $x$ from the coreset.

In this paper, we build a faster coreset generating algorithm via proposing a newly designed weighting score.
Our weighting score can be efficiently computed and therefore leads to the acceleration achieved by our algorithm.
We first introduce functions would appear in our weighting score function, e.g., approximated distance function and approximated cost function, in section~\ref{sec:approximate_distance_and_cost}.
Then we discuss the weighting score function we designed in section~\ref{sec:weighting_score}.

\subsection{Approximate Distance and Approximate Cost}\label{sec:approximate_distance_and_cost}
According to our analysis which will be shown later, we could assign weighting score based on the approximate distance function.
The approximation of the distance is again derived through the sketching technique we introduced section~\ref{sec:distance_estimation_data_structure}, which again leads to less computational load.
\begin{definition}[Approximate Distance]\label{def:approximate_distance}
Let $d_z(x,c)$ denote the $z$-th power of $\ell_2$ distance between $x \in \R^d$ and $c \in \R^d$.
We approximate $d_z(x,c)$ by $\wt{d}_z(x,c)$ where $\wt{d}_z(x,c)$ is defined as
\begin{align*}
    \wt{d}_z(x,c) := \|\Pi x - \Pi c  \|_2, 
\end{align*}
where $\Pi \in \R^{m \times d}$ is a Johnson-Lindenstrauss matrix (Lemma~\ref{lem:jl_lemma}) used for sketching. 
\end{definition}

Given the approximate distance function $\wt{d}_z(x,c)$, we define the corresponding approximate cost function $\wt{\cost}_z(U,C)$ as follows.
\begin{definition}[Approximate Cost]
    Let $\wt{d}_z(x,c)$ be an approximate distance function defined in  Definition~\ref{def:approximate_distance}.
    We define the approximate cost function with respect to $\wt{d}_z$ as
    \begin{align*}
        \wt{\cost}_z(U,C) := \sum_{x\in U}\wt{d}_z(x,C).
    \end{align*}
    When there is no ambiguity of $\wt{d}_z$, we will briefly call $\wt{\cost}_z$
    the approximate cost function.
\end{definition}

Additionally, we introduce the probabilistic projection function which will be used to build our weighting score function in the later section.
\begin{definition}[Probabilistic Projection Function]\label{def:prob_proj_func}
    Let $C^*$ be a point set in $\R^d$ represented as $C^*=\{v_1,\cdots,v_m\}$.
    Let \textsc{DE} be the sketching-based data structure for distance estimation problem which is defined in section~\ref{sec:sketching_ds_dep}.
    Suppose \textsc{DE} maintains the distance to the point set $C^*$.
    Let $t(x,z)$ denote the index reported by \textsc{DE.QueryMin}$(x,z)$
    \begin{align*}
        t(x,z):=\textsc{DE.QueryMin}(x,z)
    \end{align*}
    We define the probabilistic projection function $\wt{c}^*$ with respect to $(\textsc{DE}, C^*, z)$ as
    \begin{align*}
        \wt{c}^*(x) := v_{t(x,z)}
    \end{align*}
    When $(\textsc{DE}, C^*, z)$ is clear from the context, we will briefly call $\wt{c}^*(x)$ the probabilistic projection function.
\end{definition}

For convenience in the following steps, we also denote the preimage of the probabilistic projection function by $X_v$ for $v \in C^*$.
\begin{definition}[Preimage of Probabilistic Projection Function]\label{def:preimg_prob_proj_func}
    For $v \in C^*$ and a probabilistic projection function $\wt{c}^*:U\rightarrow C^*$, we denote the preimage set of probabilistic projection function by
    \begin{align*}
        X_v := (\wt{c}^*)^{-1}(v) = \{x \in U | \wt{c}^*(x)=v\}.
    \end{align*}
\end{definition}

\subsection{Weighting Score for Sampling}\label{sec:weighting_score}
With auxiliary functions defined in section~\ref{sec:approximate_distance_and_cost}, we introduce our weighting score function for sampling in this section.

As the algorithm we build for the coreset generation task is randomized, in addition to $U$ (the point set), $n$ (number of points), $d$ (the dimension of points), $\epsilon$ (the precision parameter for $\epsilon$-coreset), 
$(k,z)$ (parameters for the clustering problem), we also include a failuer probability parameter $\delta\in(0,0.1)$ in the input.
Given the input of the coreset generation for $(k,z)$-clustering, we first construct a $(k,\alpha)$-center set of $U$ and denote it by $C^*$. 
With the center set $C^*$, we then build the distance estimation data structure by passing $C^*$ to the sketching-based data structure.
Let \textsc{DE} be the distance estimation data structure maintaining distance to $C^*$, which was introduced in section~\ref{sec:distance_estimation_data_structure}.
Once the data structure \textsc{DE} is build, we connect points in $U$ and $C^*$ by the probabilistic projection function defined in Definition~\ref{def:prob_proj_func} and the preimage defined in Definition~\ref{def:preimg_prob_proj_func}.
Now with all the auxiliary function in section~\ref{sec:approximate_distance_and_cost} well-defined, we denote the weighting score for $x\in U$ by $\wt{\sigma}(x)$ and assign it as
\begin{align*}
    \wt{\sigma}(x) = \frac{ \wt{d}_z(x, \wt{c}^*(x))}{ \wt{\cost}_z(U,C^*)} + \frac{1}{|X_{\wt{c}^*(x)}|}
\end{align*}
for all $x\in U$.
Finally, we sample $N$ times from $U$ without replacement, where $N$ is at the order of 
\begin{align*}
    N=O(z^{O(z)} \cdot \epsilon^{-\Omega(z)} \cdot k^5 \cdot \log(k/\delta)),
\end{align*}
and in each time point $x\in U$ will be sampled with probability proportional to $\wt{\sigma}(x)$.
All the points we sampled forms the coreset as the output of our algorithm.
We present the pseudocode for our algorithm in Algorithm~\ref{alg:coreset_gen_alg}.

\subsection{Correctness and Running Time}
We formally state the correctness and running time of our coreset generation algorithm in the following two theorems.

The correctness of our algorithm includes two aspects: with the probability claimed, 1) the set $V$ output by  Algorithm~\ref{alg:coreset_gen_alg} should be an $\epsilon$-coreset for $(k,z)$-clustering problem, and 2) the size of set $V$ should be at the correct order.
Theorem~\ref{thm:main_correct:main_text} shows both aspects.
\begin{theorem}[Output Size is Nearly Linear in $k$]
	\label{thm:main_correct:main_text}
	Let $X \in \R^d$ be a given set of $n$ points. Let $\eps,\delta \in (0,0.1)$. Let $k$ be a positive integer. Let $z\geq 1$ be a real number.
	There is a randomized algorithm (Algorithm~\ref{alg:coreset_gen_alg}
	) that constructs an $\eps$-coreset $V$,   
	for $(k,z)$-clustering of size 
	\begin{align*}
	    O(\eps^{-2z-2} k \cdot \log (k) \cdot \log( k / ( \eps\delta) )),
	\end{align*}
	with probability at least $1-\delta$.
\end{theorem}
\begin{proof}
See appendix section~\ref{sec:correct_main}
\end{proof}

Theorem~\ref{thm:main_runtime:main_text} states that our algorithm runs in $\wt{O}(nk+nd)$ time, which is faster than $\wt{O}(ndk)$ in \cite{hv20} and \cite{css21}.

\begin{theorem}[Running time]
\label{thm:main_runtime:main_text}
    Let $\epsilon_0, \epsilon_1 \in (0, 0.1)$ be the constant accuracy parameter of the center set generation algorithm and the instance of distance estimation, respectively. Given an $n$-point dataset $U \in \R^d$, a coreset parameter $\epsilon \in (0,0.1)$, and a failure probability $\delta \in (0, 0.1)$, 
    Algorithm~\ref{alg:app:coreset_gen_alg} runs in time
    \begin{align*}
        &~O( (\epsilon_0^{-2} + \epsilon_1^{-2}) z^2 \cdot (nk + nd) \\
        &~ \cdot \log(k /\delta_0) \log(n/k) \log(k /\delta_1) \log(n/\delta_1)).
    \end{align*}
\end{theorem}
\begin{proof}
See appendix section~\ref{sec:main_analyze}.
\end{proof}
The framework proposed in \cite{css21} could benefit from our technique too.
Due to the space limitation, we refer to appendix section~\ref{sec:improve_cass21}.

\section{Conclusion}\label{sec:conclusion}
In this paper, we propose algorithms that can compute an $\epsilon$-coreset for $(k,z)$-clustering problem in less time, which improves previous results from $\wt{O}(ndk)$ to $\wt{O}(nd+nk)$.
As the trivial (reading data) lower bound for running time is $O(nd)$, our algorithm is very close to this lower bound.
However, it is unclear for us whether in general $O(nk)$ is also necessary.
Additionally, as our distance estimation data structure dynamically maintains distance, whether our algorithm can be adapted for the online setting is also a natural question.
We view the issue of lower bound and online setting as limitations of this paper and future work.

The implementation of our algorithm, indeed consume energy use. However, our paper provides an abstraction of large datasets efficiently, and hence reduces the further running time for later optimization procedures in machine learning pipelines. Additionally, we achieve the state-of-the-art running time. 
Thus, we hope our algorithm help reduce the energy consumption for coreset calculation, and hence any machine learning pipelines involving it.

\ifdefined\isarxivversion
\bibliographystyle{alpha}
\bibliography{ref}
\else
\bibliography{ref}
\bibliographystyle{plainnat}

\fi

\newpage
\onecolumn
\appendix
\section*{Appendix}

\paragraph{Roadmap.}

We organize the rest of our paper as follows. Section~\ref{sec:preli} gives the preliminaries for our work, introducing the notations, definitions, existing results from prior works, and other tools we use through the paper. 
Section~\ref{sec:correct} gives the formal theorem of the correctness of our coreset generation algorithm. We use a new algorithm to generate center sets, and Section~\ref{sec:center_set_gen_correct} discusses the correctness of our center set generation algorithm, and presents the proof together.
In Section~\ref{sec:improve_cass21} we present a new coreset generation algorithm generalized from the algorithm in \cite{css21}, together with its correctness and running time analysis.
Finally in Section~\ref{sec:data_structure} we present the formal version of our algorithms and data structures, including the distance estimating data structure, the center set generation algorithm and the final coreset generation algorithm, and we also give their running time analysis in this section.

\section{Preliminaries}
\label{sec:preli}

In Section~\ref{sec:notation}, we give some notations we will use throughout the paper. In Section~\ref{sec:definitions}, we introduce some definitions of our paper. In Section~\ref{sec:tri_ineq} we present the relaxed triagle inequality and its proof. In Section~\ref{sec:tail_bound_bino_dstbtn} we discussed the tail bound of binomial distributions. In Section~\ref{sec:prob_tool}, we state some probability tools we use. In Section~\ref{sec:coreset_preli}, we list some existing results on the coreset from prior works. In Section~\ref{subsec:sub_approx}, we list some existing results on the subspace approximation from previous literature.

\subsection{Notation}
\label{sec:notation}
For a given subset $A\subseteq \R^d$, we denote the convex hull of $A$ by $\mathrm{Span}(A)$. We use $[n]$ to denote the set $\{1,\ldots, n\}$. We use $\E[\cdot]$ to denote the expectation, and use $\Pr[\cdot]$ to denote the probability. We use $\|\cdot\|_{p}$ to denote the $\ell_p$ norm. For any point $x \in \R^d$ and subspace $\Gamma \subset \R^d$, we denote by $\pi_{\Gamma}(x)$ the projection of $x$ to $\Gamma$. We use $\mathcal{P}_k$ denote all subspaces of $\R^d$ with dimension at most $k$. For any function $f$, we use $\wt{O}(f)$ to denote the $O(f \cdot \poly(\log f))$. For $x, y \in \R^d$, we use $\dist(x, y)$ to denote the Euclidean distance between $x$ and $y$. 
We call a subspace $U \in \R^d$ by $k$-flat, if this subspace is of dimension $k$, i.e., $\dim (U) = k$.

\subsection{Definitions}
\label{sec:definitions}

We provide several useful definitions in this section. We begin with introducing the definition of $(k,z)$-clustering problem, where the goal is to find a collection of  $k$ points, such that the total cost is minimized.

\begin{definition}[$(k ,z)$-Clustering Problem]
\label{def:kz_clustering}
    The goal of the $(k,z)$-Clustering problem is, for a given $n$-point set $U \subseteq \R^d$, to find a $k$-point set $C \subseteq \R^d$, called a \emph{center set} that minimizes the cost function
    \begin{align*}
        \cost_z(U, C) := \sum_{x \in U} d_z(x ,C),
    \end{align*}
    where the $d_z(x, C) \in \R_{\geq 0}$ is defined as
    \begin{align*}
        d_z(x, C) := \min_{c \in C} \|x - c\|_2^z.
    \end{align*}
\end{definition}

Next, we present the definition of $(k, \alpha)$-center set, which serves as an approximation benchmark for the $(k,z)$-clustering problem.

\begin{definition}[$(k, \alpha)$-Center Set]
\label{def:alhpa_center_set}
    Given $\alpha \in [1, +\infty)$ 
    and a set $U \in \R^d$ of $n$ points, we call a set $V^* \subset \R^d$ a $(k, \alpha)$-center set 
    for a given $U \subset \R^d$ 
    if it satisfies
    \begin{align*}
        \cost_z(U, V^*) \le \alpha \cdot \cost_z(U, V)
    \end{align*}
    for any other set $V$ with $k$ points.
\end{definition}

Furthermore, we present the definition of a ball of radius $r$ for a point $x$ and a set $U$.

\begin{definition}[Ball]\label{def:ball}
    For any $x \in \R^d$, we define the ball $B(x,r) \subset \R^d$ centered at x with radius $r$ as
    \begin{align*}
        B(x, r) = \{y \in \R^d ~|~ d_z(x, y) \le r \},
    \end{align*}
    and for a set $U$, we define $B(U,r)$ as the union of the balls $B(x,r)$ for all $x \in U$:
    \begin{align*}
        B(U, r) = \bigcup_{x \in U}B(x, r).
    \end{align*}
\end{definition}

Additionally, we present the definition for $(k,z)$-subspace approximation problem, whose goal is finding a subspace $P$ minimizing the total projection loss of $U$ to $P$. 

\begin{definition}[$(k,z)$-Subspace Approximation]
\label{def:subspace_prob}
     The objective of the $(k, z)$-subspace approximation problem is that, given a positive integer $k$, a constant $z>0$, and a dataset $U \in \R^d$,
     find a subspace $P \in \mathcal{P}_k$ that minimizes the sum of distances, i.e.,
    \begin{align*}
        \sum_{x \in U}d_z(x, P) .
    \end{align*}
\end{definition}

Here we present the definition of ``weak coreset'' for the subspace approximation problem as follows.

\begin{definition}[Weak-coreset for $(k,z)$-Subspace Approximation]
\label{def:weak_coreset}
    For a given set $U \subseteq \R^d$ of $n$ points. Let $\epsilon \in (0, 0.1)$. We say a subset $S \in \R^d$ together with a weight function $w:S \rightarrow \R_{\ge 0}$ is an $\epsilon$-weak-coreset for the $(k, z)$-subspace approximation (Definition~\ref{def:subspace_prob}) if
    \begin{align*}
        \min_{P \in \mathcal{P}_k}\sum_{x \in S}w(x) \cdot d_z(x, P) \in (1 \pm \epsilon) \cdot \min_{P \in \mathcal{P}_k}\sum_{x \in U}d_z(x, P).
    \end{align*}

\end{definition}

Moreover, we define the set of all sets with size $k$ as follows:

\begin{definition}[Domain of Sets with Size $k$]
\label{def:all_center_set}
    We define the set of all sets with size $k$ in $\R^d$ as
    \begin{align*}
        \mathcal{C}_k := \{C = (c_1, \dots, c_k) ~|~ c_i \in \R^d, \forall i \in [k]\}.
    \end{align*}
    For simplicity of notation, we use $\mathcal{C}$ to denote $\mathcal{C}_k$.
\end{definition}

Therefore, we can formally introduce the goal of the $(k,z)$-clustering problem: To find a set $C^* \in {\cal C}$ that approximately minimizes the total cost of set $U$ on $C^*$.

\begin{definition}[$\eps$-Approximate Center]
\label{def:approximate_center_set}
    Given a set $U$ of $n$ points in $\R^d$, we denote a center set $C^*$ is an $\epsilon$-approximate solution for $(k, z)$-clustering problem of the set $U$ if
    \begin{align*}
        \min_{C \in \mathcal{C}} \cost_z(U, C) \in (1 \pm \epsilon)\cost_z(U, C^*)
    \end{align*}
\end{definition}

On top of that, we present the definition of equivalence relations and classes for sets of size $k$ based on the projections onto subspaces. 
\begin{definition}[Relations of Equivalence]
	\label{def:equivalent}
	
	For a given subspace $\Gamma\subsetneq \R^d$, we define an relation $\sim_\Gamma$, called equivalent class, as follows: for two sets
	$C= \{c_1,\ldots,c_k \}, \wt{C} = \{ \wt{c}_1, \ldots, \wt{c}_k\} \in {\cal C}$(Definition~\ref{def:all_center_set}), we say there is equivalence relation $\sim_\Gamma$ between $C$ and $\wt{C}$ with respect to subspace $\Gamma$, if for all $i\in [k]$,
	\begin{align*}
    	\pi_\Gamma(c_i) = \pi_\Gamma( \wt{c}_i) \quad \text{and} \quad d(c_i,\pi_\Gamma(c_i)) = d( \wt{c}_i,\pi_\Gamma( \wt{c}_i)),
	\end{align*}
	where for any $x \in \R^d$, $\pi_{\Pi}(x)$ denotes its projection on $\Pi$.
	We define $\wt{\Gamma}$ to be such a subspace which is obtained from $\Gamma$ by appending an arbitrary dimension $u \in \R^d$, i.e.,  
	\begin{align*} 
	\wt{\Gamma} := \left\{ a \cdot x+b \cdot u \mid x\in \Gamma, a\in \R,b\in \R\right\},
	\end{align*}
	where $u$ is a point on a arbitrary one-dimensional subspace in $\R^d$ that is orthogonal to $\Gamma$.
	Additionally, we use $\mathcal{C}_\Gamma$ to denote the set of all sets $C \in \mathcal{C}$ with size $k$ whose points all lie in $\wt{\Gamma}$, i.e.,
	\begin{align*}
	\mathcal{C}_\Gamma:= \{C=\{c_1,\ldots,c_k\} \in \mathcal{C} ~|~ c_i\in  \wt{\Gamma}, ~ \forall i\in[k]\}.
	\end{align*}
	With relation $\sim_\Gamma$, we define equivalence classes of sets with size $k$, i.e., $\{\Delta^{\Gamma}_C: C\in \mathcal{C}_\Gamma \}$, where each \begin{align*}
	\Delta^{\Gamma}_C :=  \{ \wt{C} \in \mathcal{C}: C\sim_\Gamma \wt{C} \}.
	\end{align*}
\end{definition}

Finally, we present the definition of $\epsilon$-representativeness property of a weighted point set.

\begin{definition}[Representativeness Property]
\label{def:represent_property}
    For a given weighted set $U \subset \R^d$ and a weight function $w: U \rightarrow \R_+$. Given a subspace $\Gamma$. Given $\epsilon \in (0, 0.1)$, we say $U$ satisfies the $\epsilon$-representativeness property with respect to $\Gamma$ if for any equivalence class $\Delta_C^{\Gamma}$ and any two sets $C_1, C_2 \in \Delta_C^{\Gamma}$, the following property holds:
    \begin{align*}
        \cost_z(U, C_1) \in (1 \pm \eps) \cdot \cost_z(U, C_2).
    \end{align*}
\end{definition}

\subsection{Relaxed Triangle Inequality }
\label{sec:tri_ineq}

Here we present the relaxed triangle inequality for $d_z$ associate with its proof as follows.
\begin{lemma}[Relaxed Triangle Inequality for $d_z$]
\label{lem:relax_tri_ineq}
    For any $x_1, x_2, x_3 \in \R^d$, we have
    \begin{align*}
        d_z(x_1, x_3) \le 2^z \cdot (d_z(x_1, x_2) + d_z(x_2, x_3)). 
    \end{align*}
\end{lemma}

\begin{proof}
    We have 
    \begin{align*}
        d_z(x_1, x_3)
    =   & ~ \|x_1 - x_3\|_2^z \\
    \le & ~ (\|x_1 - x_2\|_2 + \|x_2 - x_3\|_2)^z \\
    = & ~ 2^z \cdot (\frac{1}{2}\|x_1 - x_2\|_2 + \frac{1}{2}\|x_2 - x_3\|_2)^z \\
    \le & ~ 2^z \cdot ( \frac{1}{2} \|x_1 - x_2\|_2^z + \frac{1}{2} \|x_2 - x_3\|_2^z ) \\
    \le & ~ 2^z \cdot ( \|x_1 - x_2\|_2^z + \|x_2 - x_3\|_2^z ) \\
    =   & ~ 2^z \cdot (d_z(x_1, x_2) + d_z(x_2, x_3)), 
    \end{align*}
    where the first step follows from the definition of $d_z$, the second step follows from the triangle inequality, the third step follows from taking out the factor $2^z$, the fourth step follows from $f(x) = x^z$ is convex when $z \ge 1$ and $x \ge 0$, the fifth step follows from $1/2 \le 1$, and the last step follows from the definition of $d_z$.
    
    Thus we complete the proof.
\end{proof}

\subsection{Tail Bounds for the Binomial Distribution}
\label{sec:tail_bound_bino_dstbtn}

In this section we state several standard bounds on the tail of the binomial distribution. Let $n$ be a non-negative integer and let $p$ be a real in $[0, 1]$. Let $X$ denote the random variable corresponding to the total number of successes in $n$ independent Bernoulli trials, each of which succeeds with probability $p$. The random variable $X$ is said to be \emph{binomially distributed} with parameters $n$ and $p$. Note that $\E[X] = n p$. Let $\mu$ denote $\E[X]$. Then, we can use Chernoff bound (Lemm~\ref{lem:chernoff_bound}).

\subsection{Probability Tools}
\label{sec:prob_tool}

Then we introduce some useful probability tools. We begin by presenting the Chernoff bound, which upper bounds the probability that the sum of binary variables deviates from its mean. 

\begin{lemma}[Chernoff Bound \cite{che52}]\label{lem:chernoff_bound}
    Let $X = \sum_{i=1}^n X_i$, where $X_i = 1$ with probability $p_i$ and $X_i = 0$ with probability $1 - p_i$, and all $X_i$ are independent. Let $\mu = \E[X] = \sum_{i = 1}^n p_i$. Then
    \begin{itemize}
        \item $\Pr[X \ge (1+\delta)\mu] \le \exp(-\delta^2\mu/3)$, $\forall \delta > 0$;
        \item $\Pr[X \le (1-\delta)\mu] \le \exp(-\delta^2\mu/2)$, $\forall 0 < \delta < 1$.
    \end{itemize}
\end{lemma}

Next, we present the Hoeffding bound, which upper bound the probability that the sum of independent bounded variables deviates from its mean. 

\begin{lemma}[Hoeffding Bound \cite{hoe63}]
\label{lem:hoeffding}
    Let $X_1, \dots, X_n$ denote $n$ independent bounded variables in $[a_i, b_i]$. Let $X= \sum_{i=1}^n X_i$, then we have
    \begin{align*}
        \Pr[|X - \E[X]| \ge t] \le 2 \exp(-\frac{2t^2}{\sum_{i=1}^n(b_i-a_i)^2}).
    \end{align*}
\end{lemma}

Finally, we present the JL-lemma, which states that there exists a projection mapping the high-dimensional points to a lower dimensional space, which maintains the distance of these points.

\begin{lemma}[Johnson–Lindenstrauss Lemma, \cite{jl84}]
\label{lem:jl_lemma}
    Let $\Pi$ in $\R^{m \times d}$ denote a random Gaussian matrix, if $m = \epsilon^{-2} \log (n/\delta)$, then for any fixed set of points $S$ with $|S|=n$, we have with probability $1 - \delta$,
    \begin{align*}
        (1 - \eps)\| x - y \|_2 \le \| \Pi x - \Pi y \|_2 \le (1 + \eps)\| x - y \|_2
    \end{align*}
    for all $x, y$ in $S$.
\end{lemma}

\subsection{Coresets}
\label{sec:coreset_preli}

In this section, we present some existing results on coresets that we will use in the remainder of the paper. We begin by introducing the two framework by importance sampling . 

\begin{theorem}[Feldman-Langberg Framework~\cite{fl11, bfl+16}]
	\label{thm:fl11_bfl16}
	Given a weighted set $X \in \R^d$ with $n$ points, let $u: X \rightarrow \R_{\geq 0}$ denote its weight function.
	Let parameters $\eps \in (0, 0.1)$.
	Let constant $z\geq 1$ and $k\geq 1$.
	Let $C^*\in \mathcal{C}_k$ (Definition~\ref{def:all_center_set}) 
	denote a center set with size of $k$ 
	that is an $O(1)$-approximate solution for $(k, z)$-Clustering of the set $X$ (Definition~\ref{def:approximate_center_set}).
	We have following two frameworks of importance sampling.
	\begin{enumerate}
	\item (Theorem~15.5 in \cite{fl11})  Let $\epsilon \in (0,0.1)$ denote an accuracy parameter. Let $\delta \in (0,0.1)$ denote the failure probability. 
	Let $\sigma:X\rightarrow \R_{\geq 0}$ be such a function satisfying,
	\begin{align*}
	    \sigma(x) \geq \frac{ u(x)\cdot d_z(x,C^*)}{\sum_{y\in X} u(y)\cdot d_z(y,C^*)},
	\end{align*}
	for any $x\in X$. We define $\G := \sum_{x \in X} \sigma(x)$.
	Let $D\subseteq X$ be a set of points generated by taking 
	\begin{align*}
	    O\left(\eps^{-2z} ( d k \log k+\log(1/\delta))\right)
	\end{align*}
	samples from $X$, where each sample $x\in X$ is selected with probability $\frac{\sigma(x)}{\G}$ and let its weight be $w(x):= \frac{\G}{|D|\cdot \sigma(x)}$.
	For every $c\in C^*$
	, define $w_{C^*}(c):= (1+10\eps)\cdot\sum_{x\in X_c}u(x)-\sum_{x\in D\cap X_c} w(x)$ where $X_c$ is defined as 
	\begin{align*}
	    X_c := \{y \in X ~|~ c = \arg\min_{z \in C^*}\mathrm{dist}(y, z)\}.
	\end{align*}
    Define $S := D\cup C^*$. Then by the construction above, we have that, $S$ is an $\epsilon$-coreset for $(k, z)$-clustering problem over the original set $X$, with probability at least $1 - \delta$.
	\item (Theorem~5.2 in \cite{bfl+16}) 
    Let $\epsilon \in (0,0.1)$ denote an accuracy parameter. Let $\delta \in (0,0.1)$ denote the failure probability. 
    Let $\sigma:X\rightarrow \R_{\geq 0}$ be a function such that, 
	\begin{align*}
	    \sigma(x) \geq \sup_{C\in \mathcal{C}} \frac{u(x)\cdot d_z(x,C)}{\sum_{y\in X} u(y)\cdot d_z(y,C)}
	\end{align*}
	 for any $x\in X$, and we define 
	 $\G := \sum_{x\in X}\sigma(x)$.
	Let $S\subseteq X$ be a set of points generated by taking 
	\begin{align*}
	    O\left(\eps^{-2} \G (dk \log (\G) +\log(1/\delta))\right)
	\end{align*}
	samples from $X$, where each sample $x\in X$ is selected with probability $\frac{\sigma(x)}{\G}$ and let its weight be $w(x):= \frac{\G}{|S|\cdot \sigma(x)}$.
    Then, we have that, $S$ is an $\eps$-coreset for $(k, z)$-clustering over $X$ with probability at least $1-\delta$.
	\end{enumerate}
\end{theorem}

The following lemma proved the existence of a subspace $\Gamma$, such that all $(k,z)$-clustering objectives is well-estimated by the projections of $X$ onto $\Gamma$. \cite{sw18} studies the unweighted point sets, it is not hard to see that their results can be extended to the weighted point sets. 

\begin{lemma}[Lemma 6 and Theorem 10 in \cite{sw18}]
	\label{lem:projection}
	For a given weighted set $A\subseteq \R^d$, let $w:A \rightarrow \R_{\geq 0}$ be its weight function. Let constant $z \geq 1$. Denote the optimal objective for weighted $(k,z)$-clustering over $A$ by $\OPT_z$. Let $\Gamma \in \R^d$ be a subspace. Let $\eps \in (0, 0.1)$, we define $\epsilon_z :=\frac{ \eps^{z+3}}{3\cdot (100z)^{2z}}$. Assume for any set $C\in \mathcal{C}_k$, we have 
	\begin{align}
	\label{eq:projection1}
    	\sum_{x\in A} w(x)\cdot (d_z(x,\pi(x))-d_z(x,\pi_C(x)))\leq \epsilon_z \cdot \OPT_z,
	\end{align}
	where for any $x \in A$, let $\pi(x)$ be the projection of $x$ onto $\Gamma$, and $\pi_C(x)$ be its projection onto $\mathrm{Span}(\Gamma\cup C)$, and $d_z$ is defined in Definition~\ref{def:kz_clustering}.
	Then for any set $C\in \mathcal{C}_k$, we have{\small
	\begin{align}
    	\label{eq:projection2}
    	    (1 - \eps)\cdot \sum_{x\in A} w(x)\cdot d_z(x,C) 
    	\leq  \sum_{x\in A} w(x)\cdot (d_2(\pi(x),C)+d_2(x,\pi(x)))^{z/2}
    	\leq  (1 + \eps)\cdot \sum_{x\in A} w(x)\cdot d_z(x,C).
	\end{align}}
\end{lemma}

\begin{lemma}[Lemma~6 in \cite{sw18}]
\label{lem:subspace_guarant}
    Let $X \in \R^d$ denote a set of $n$ points. There exists an $O(k/\epsilon_z)$-dimensional subspace $\Gamma$ satisfying both $C^* \subseteq \Gamma$, and that
    \begin{align*}
        \sum_{x\in X} (d_z(x,\pi(x))-d_z(x,\pi_C(x)))\leq \epsilon_z \cdot \OPT_z/2,
    \end{align*}
     for any $C \in \mathcal{C}$, where for any $x \in X$, define $\pi(x) := \arg\min_{y \in \Gamma} \mathrm{dist}(x, y)$ to be the projection of $x$ to $\Gamma$, and define the projection of $x$ to $\mathrm{Span}(\Gamma\cup C)$ by $\pi_C(x)$. 
    
\end{lemma}

\subsection{Subspace Approximation}\label{subsec:sub_approx}

In this section, we present two results from previous literature on subspace approximation. The first result is a guarantee on the set of an importance sampling method. The second results guarantees that there are approximate $k$-flats in low dimensional subspaces.

\begin{lemma}[Weak-Coreset for Subspace Approximation, Theorem~5.10 in \cite{hv20}]
	\label{lem:subspace}
	Let $X \in \R^d$ be a set of $n$ points. Let $\eps \in (0,0.1)$ denote an accuracy parameter. Let $\delta \in (0,0.1)$ denote the failure probability. Let $z \geq 1$ be constant. Let $k$ denote a positive integer. Assume that $\sigma_0: X \rightarrow \R_{\geq 0}$ is a sensitivity function satisfying
	\begin{align*}
    	\sigma_0(x) \geq \sup_{P\subseteq \mathcal{P}_k} \frac{d_z(x, P)}{\sum_{y\in X} d_z(y, P)} , ~~~ \forall x \in X.
	\end{align*} 
	We define $\G := \sum_{x\in X} \sigma_0(x)$. We define $w(x):= \frac{\G}{|S|\cdot \sigma_0(x)}$. Let $S\subseteq X$ be a set of points which is constructed by taking
	\begin{align*}
	    O( \eps^{-2}  \G^2 \cdot (\eps^{-1} k^3\log (k/\eps)+\log(1/\delta)))
	\end{align*} 
	samples. In the sampling procedure, each sample $x\in X$ is chosen with probability $\frac{\sigma_0(x)}{\G}$. For each $x$, we set weight function to be $w(x)$.
	
	Then we have, with probability at least $1-\delta$, $S$ is an $\eps$-weak-coreset (Definition~\ref{def:weak_coreset})
	 for the $(k,z)$-subspace approximation problem of the set $X$.
\end{lemma}

\begin{lemma}[There exist an approximate $k$-flats in low dimensional subspaces, Lemma~5.11 in \cite{hv20}]
	\label{lem:projective_clustering}
	For a given weighted set of points $X \in \R^d$ of $n$ points, let $u(x): X \rightarrow \R^{\ge 0}$ be its weight function. Let $ \eps \in (0,0.1)$ be an accuracy parameter.  Let $z\geq 1$ be constant. Let  $k \geq 1$ be an integer. Let $K:=O(\eps^{-1} k^2 \log(k/\eps))$.  Then, there is a $k$-flat 
	$P$ which is spanned by at most $K$ points in $X$, satisfying
	\begin{align*}
    	\sum_{x\in X} u(x)\cdot d_z(x,P)\leq (1+\eps)\cdot \min_{P'\in \mathcal{P}_k}\sum_{x\in X} u(x)\cdot d_z(x,P').
	\end{align*}
\end{lemma}

\section{Correctness}
\label{sec:correct}

In this section, we state and prove the correctness of Algorithm~\ref{alg:coreset_gen_alg}. In Section~\ref{sec:correct_def}, we present additional definitions used in the proof. 
In Section~\ref{subsec:exact_metrics}, we present the exact metrics, and then in Section~\ref{subsec:approx_metrics}, we present the approximate metrics. In Section~\ref{sec:correct_main}, we state the main correctness theorem associated with its proof. In Section~\ref{sec:correct_repre_prop}, we discuss the representiveness property of the original set and the output set. In Section~\ref{sec:correct_bound_proportion}, we bound the proportion of cost of each point. In Section~\ref{sec:correct_output_is_coreset}, we state that the output set is a coreset in a low-dimensional space. In Section~\ref{subsec:approx_distance}, we present the approximate distance preserving lemma. 
In Section~\ref{sec:correct_preserv_prop}, we discuss the preservative property of subspace. 
In Section~\ref{sec:weight_dist_close_to_uniform}, we show the weighted sum of distances is lower bounded by the uniform sum.

\subsection{Definitions}
\label{sec:correct_def}

In this section, we define more variables that we will use in the remainder of the paper. At first, we state the definition of function $\epsilon_z$.

\begin{definition}[$\epsilon_z$]\label{def:epsilon_z}
We define $\epsilon_z \in (0,0.1)$ as follows
\begin{align*}
    \epsilon_z :=  \frac{\epsilon^{z+3}}{10\cdot (100 z)^{2z}}
\end{align*}
\end{definition}
It is obvious that we can upper bound $\epsilon_z$ by $O(\epsilon)$ if $z \geq 0.1$ by the following claim:
\begin{claim}[Upper bound of $\epsilon_z$]\label{cla:upper_bound_epsilon_z_by_epsilon}
        For any constant $z \geq 0.1$, we have
        \begin{align*}
            \epsilon_z \leq O(\epsilon).
        \end{align*}
\end{claim}

Additionally, we use $C^*$ to denote a $(k,\gamma)$-center set:

\begin{definition}[$C^*$]
\label{def:C_star}
    Let $\gamma=O(1)$ and $\gamma \ge 1$. Let $C^*$ denote a set in $\mathcal{C}$ such that it is an $(k, \gamma)$-center set (Definition~\ref{def:alhpa_center_set}).
\end{definition}

\subsection{Exact Metrics}\label{subsec:exact_metrics}

In this section, we present some definitions on the exact metric.
Given a set $X$ and a point $x \in X$, we define $c^*(x)$, $X_c$, $\sigma(x)$ as follows.
\begin{definition}
\label{def:c_func}
    Let $X$ be a set in $\R^d$. Let $C$ be defined as Definition~\ref{def:C_star}. Then we define
    \begin{align*}
        c^*(x) := \arg\min_{c \in C^*} d_z(x, c)
    \end{align*}
    for any $x \in X$.
\end{definition}

\begin{definition}
\label{def:cluser_set}
    Let $X$ be a set in $\R^d$. Let $c^*(x)$ be defined as Definition~\ref{def:c_func}. 
    Then, for any $c \in C^*$, we define $X_c \subset \R^d$
    \begin{align*}
        X_c := \{y \in X ~|~ c^*(y) = c\}.
    \end{align*}
\end{definition}

\begin{definition}
\label{def:sigma_func}
    Let $c_0 = 4$ be a fixed constant. 
     Let $X$ denote a fixed set in $\R^d$. Let $C^*$ be defined as Definition~\ref{def:C_star}. Let $X_c$ be defined as Definition~\ref{def:cluser_set}.
    Define
    \begin{align*}
        \sigma(x) := c_0 \cdot 2^{2z } \gamma^2 \cdot \big( \frac{d_z(x,c^*(x))}{\cost(X,C^*)}+\frac{1}{|X_{c^*(x)}|} \big),
    \end{align*}
    for any $x \in X$.
\end{definition}

\subsection{Approximate Metrics}\label{subsec:approx_metrics}
In this section, we present the definition of approximate metrics, i.e., $\tilde{c}^*$ and $\tilde{\delta}$. 
Let $\wt{d}_z$ denote a fixed metric such that it approximates $d_z$ (for details, we refer the readers to Lemma~\ref{lem:approximate_cost}). 

\begin{definition}
\label{def:c_func_approx}
    Let $X$ be a set in $\R^d$. Let $C$ be defined as Definition~\ref{def:C_star}. Then we define
    \begin{align*}
        \wt{c}^*(x) := \arg\min_{c \in C^*} \wt{d}_z(x, c)
    \end{align*}
    for any $x \in X$.
\end{definition}

\begin{definition}
\label{def:cluser_set_approx}
    Let $X$ be a set in $\R^d$. Let $\tilde{c}^*(x)$ be defined as Definition~\ref{def:c_func_approx}. 
    Then, for any $c \in C^*$, we define $X_{\tilde{c}} \subset \R^d$
    \begin{align*}
        X_{\tilde{c}} := \{y \in X ~|~ \tilde{c}^*(y) = c\}.
    \end{align*}
\end{definition}

\begin{definition}
\label{def:sigma_func_approx}
    
    Let $\wt{c}^*$ be defined as Definition~\ref{def:c_func_approx}. Let $\wt{c}_0 = 16$ be a fixed constant. 
     Let $X$ denote a fixed set in $\R^d$. Let $C^*$ be defined as Definition~\ref{def:C_star}. Let $\wt{\cost}_z(X,C^*) := \sum_{x \in X} \wt{d}_z(x,C^*)$ Let $X_c$ be defined as Definition~\ref{def:cluser_set}.
    Define
    \begin{align*}
        \wt{\sigma}(x) := \wt{c}_0 \cdot 2^{2z} \gamma^2 \cdot \big( \frac{ \wt{d}_z(x, \wt{c}^*(x))}{ \wt{\cost}_z (X,C^*)}+\frac{1}{|X_{ \wt{c}^*(x)}|} \big),
    \end{align*}
    for any $x \in X$.
\end{definition}

\subsection{Main Result}
\label{sec:correct_main}
In this section, we present the correctness theorem(Theorem~\ref{thm:main_correct}) of our proposed algorithm, together with its proof.

\begin{definition}[Optimal Objective]
\label{def:opt_cost}
    Let $X\subseteq \R^d$ be a given set of $n$ points. Let $\mathcal{C}_k$ be defined as in Definition~\ref{def:all_center_set}. For a set of points $U \in \R^d$, let $\OPT_z$ be the optimal objective of the  $(k,z)$-clustering problem of the original set $X$, i.e.,
    \begin{align*}
        \OPT_z := \min_{C \in \mathcal{C}_k} \cost_z(U, C).
    \end{align*}

\end{definition}

We provide the correctness result for Coreset Generation as follows,
\begin{theorem}[Output Size is Nearly Linear in $k$]
	\label{thm:main_correct}
	Let $X \in \R^d$ be a given set of $n$ points. Let $\eps,\delta \in (0,0.1)$. Let integer $k \geq 1$. Let constant $z\geq 1$. There is a randomized algorithm (Algorithm~\ref{alg:coreset_gen_alg}
	) that constructs an $\eps$-coreset $D$,   
	for $(k,z)$-clustering of size 
	\begin{align*}
	    O(\eps^{-2z-2} k \cdot \log (k) \cdot \log( k / ( \eps\delta) )),
	\end{align*}
	with probability at least $1-\delta$.
\end{theorem}

\begin{proof}

    In Algorithm~\ref{alg:coreset_gen_alg}, we first run \textsc{CenterSetGen}(Algorithm~\ref{alg:center_set_gen}) and obtain $C^*$. 
    By Theorem~\ref{thm:center_set_gen_correct}, we know $C^*$ satisfies that, with high probability $1 - e^{c_0 k}$, for any $C \in \mathcal{C}_k$, 
    \begin{align*}
        \cost_z(X, C^*) \le \gamma \cdot \cost_z(X, C).
    \end{align*}
    
    Next, we run the remaining steps of Algorithm~\ref{alg:coreset_gen_alg}. We analyze the correctness for the remaining steps in the next few paragraphs.

    Let $\epsilon_z$ be defined as Definition~\ref{def:epsilon_z}. Then by Lemma~\ref{lem:subspace_guarant}, there exists an $O(k/\epsilon_z)$-dimensional subspace $\Gamma$ satisfying both $C^* \subseteq \Gamma$, and that 
    \begin{align*}
        \sum_{x\in X} (d_z(x,\pi(x))-d_z(x,\pi_C(x)))\leq \epsilon_z \cdot \OPT_z/2,
    \end{align*}
     for any $C \in \mathcal{C}$, where for any $x \in X$, define $\pi(x) := \arg\min_{y \in \Gamma} \mathrm{dist}(x, y)$ 
      to be the projection of $x$ to $\Gamma$, and define the projection of $x$ to $\mathrm{Span}(\Gamma\cup C)$ by $\pi_C(x)$.

    Next, we append $\Gamma$ by an arbitrary dimension in $\R^d$, and denote it as $\Gamma'$. Note that the dimension of $\Gamma$ is $O(k/\epsilon_z)$, we deduce that $\Gamma'$ is also $O(k/\epsilon_z )$-dimensional. Let $D$ be the output set of Algorithm~\ref{alg:coreset_gen_alg}. Then, note the way we construct $D$, by Lemma~\ref{lem:vx12}, we get that, $D$ is an $\eps$-coreset for $(k, z)$-Clustering in $\Gamma'$,
     with probability at least $1-\delta/10$, which means that for any set $C\subset \Gamma'$ with size $k$,
	\begin{align}
	\label{eq:Q2}
	    \sum_{x\in D} u(x)\cdot d_z(x,C) \in (1\pm \eps)\cdot \cost_z(X,C).
	\end{align}
	
	Then, using Eq.~\eqref{eq:projection2} in Lemma~\ref{lem:projection}, for any set $C\in \mathcal{C}$, we have:
	\begin{align}
	\label{eq:Q1} 
	    \sum_{x\in X} (d_2(\pi(x),C)+d_2(x,\pi(x)))^{z/2}\in (1\pm \eps)\cdot \cost_z(X,C)
	\end{align}
	with probability at least $1-\delta/10$.
	
	Next, by Lemma~\ref{lem:preserve}, we have that: 
	\begin{align}
	\label{eq:coreset_preserve}
	    \sum_{x\in D} u(x)\cdot d_z(x,C)\in (1\pm 2 \eps) \sum_{x\in D} u(x)\cdot (d_2(\pi(x),C)+d_2(x,\pi(x)))^{z/2}.
	\end{align}
	
	Then, for a given set $C\in \mathcal{C}$, suppose $C$ is a member of the equivalence class $\Delta^\Gamma_{C'}$ (Definition~\ref{def:equivalent}) for some $C'\in \Gamma'$. Here we note that, the dimension of $\Gamma$ is $O(k/\epsilon_z)$, which we assume is $< d$. Thus we have this $C'$ is guaranteed to exist. 
	We have that:
	
	\begin{align*}
			    & ~ \sum_{x\in D} u(x)\cdot d_z(x,C) \\
			\in & ~ (1\pm 2\eps)\cdot \sum_{x\in D} u(x)\cdot d_z(x,C') \\
			\in & ~ (1\pm 2\eps) \cdot \cost_z(X,C') \\
			\in & ~ (1\pm 4\eps) \cdot \cost_z(X,C),
	\end{align*}

	where the first step follows from Claim~\ref{cla:alg}, the second step follows from Eq.~\eqref{eq:Q2}, the third step follows again from Claim~\ref{cla:alg}.
	
	Thus we complete the proof.
\end{proof}

\subsection{Representativeness Property of Original Set and Output Set}
\label{sec:correct_repre_prop}
In this section, we present the proof on the $(2\epsilon)$-representativeness of both set $X$ and set $D$.

	\begin{claim}
		\label{cla:alg}
		Let $\Gamma$ be a subspace satisfying Lemma~\ref{lem:subspace_guarant} 
		and setting $\epsilon_z$ as in Definition~\ref{def:epsilon_z}, we have that, the original set $X$ and the output set $D$ of the Algorithm~\ref{alg:coreset_gen_alg} 
		both satisfy the $(2\eps)$-representativeness property (Definition~\ref{def:represent_property}) with respect to $\Gamma$.
	\end{claim}
	
	\begin{proof}
		Let $\Delta_C^{\Gamma}$ be an arbitrary equivalence class on subspace $\Gamma$ (Definition~\ref{def:equivalent}). 
		Let $C_1,C_2\in \Delta_C^{\Gamma}$ be two arbitrary set with size of $k$, we have
		\begin{align*}
				     \cost_z(X,C_1) 
			    \in & ~ (1\pm \eps)	\cdot \sum_{x\in X} (d_2(\pi(x),C_1)+d_2(x,\pi(x)))^{z/2} \\
				\in & ~ (1\pm \eps) \cdot \sum_{x\in X} (d_2(\pi(x),C_2)+d_2(x,\pi(x)))^{z/2} \\
				\in & ~ (1\pm 2\eps)\cdot \cost_z(X,C_2),
		\end{align*}
		where the first step follows from Eq.~\eqref{eq:Q1}, the second step follows from the definition of equivalence class(Definition~\ref{def:equivalent}), and the third step follows from Eq.~\eqref{eq:Q1}.

		In a similar manner, we can show by Eq.~\eqref{eq:coreset_preserve}, the set $D$ meets the $2 \eps$-representativeness 
		property as well.
		
		Hence, we complete the proof.
	\end{proof}

\subsection{Bounding the Proportion of Cost}
\label{sec:correct_bound_proportion}
In this section, we present the proof of the upper bound on the quantity $\frac{d_z(x,C)}{\cost_z(X,C)}$. Note this claim gives the same bound as in Claim~5.6 in previous work \cite{hv20}. However, our claim is different from the previous one in that we use \emph{approximate} metrics. In this claim, we show that, since our approximate distance metric is still close to the exact ones, we get the following bound.

\begin{claim}
		\label{cla:sen_bound}
		Let $\mathcal{C}$ be defined as Definition~\ref{def:all_center_set}. Let $\cost_z(X, C)$ be defined as Definition~\ref{def:kz_clustering}. Let $\OPT_z$ be defined as Definition~\ref{def:opt_cost}. Let   
		$c^*$ and $X_{c^*(x)}$ be defined as Definition~\ref{def:cluser_set}. Let $\wt{c}^*(x)$ be defined in Definition~\ref{def:c_func_approx}. Then, we have
		\begin{align*}
    		\sup_{C\in \mathcal{C}} \frac{d_z(x,C)}{\cost_z(X,C)}\leq 8 \cdot ( 2^{z} \cdot \frac{d_z(x,c^*(x))}{\OPT_z}+ 2^{2z}\gamma\cdot \frac{1}{|X_{\wt{c}^*(x)}|} ).
		\end{align*}

	\end{claim} 
    \begin{proof}
	Fix a point $x\in X$ and a set $C\in \mathcal{C}$, we denote the nearest point to $x$ in $C$ to be $y$, i.e., $y := \arg\min_{y \in C}d_z(x, y)$ then we have
	\begin{align}
	\label{eq:sen1}
	        d_z(x, C)
	    =   & ~ d_z(x, y) \notag\\
	    \leq & ~ 2^z \cdot (d_z(x,\wt{c}^*(x)) + d_z( \wt{c}^*(x),y) ) \notag \\
	    \leq & ~ 2^z \cdot (d_z(x,\wt{c}^*(x)) + d_z( \wt{c}^*(x),C) ) \notag \\
	    \leq & ~ 2^{z+1} \cdot \wt{d}_z(x, \wt{c}^*(x)) + 2^{z+1}  \cdot \wt{d}_z( \wt{c}^*(x),C) ) \notag \\
    	\leq & ~ 2^{z+1} \cdot \wt{d}_z(x, \wt{c}^*(x)) + \frac{2^{z+1} }{|X_{\wt{c}^*(x)}|}\cdot \sum_{y\in X} \wt{d}_z( \wt{c}^*(y),C) \notag \\ 
	    \leq & ~ 2^{z+1} \cdot \wt{d}_z(x, \wt{c}^*(x)) + \frac{2^{z+1} }{|X_{\wt{c}^*(x)}|}\cdot \sum_{y\in X} 2^z\cdot ( \wt{d}_z( \wt{c}^*(x),x)+ \wt{d}_z(x,C)) \notag \\
	    \leq & ~ 2^{z+1} \cdot \wt{d}_z(x, c^*(x)) + \frac{2^{z+1} }{|X_{\wt{c}^*(x)}|}\cdot \sum_{y\in X} 2^z\cdot ( \wt{d}_z( c^*(x),x)+ \wt{d}_z(x,C)) \notag \\
	    \leq & ~ 2^{z+1} \cdot \wt{d}_z(x, c^*(x)) + \frac{2^{z+1} }{|X_{\wt{c}^*(x)}|}\cdot \sum_{y\in X} 2^z\cdot ( 2 d_z( c^*(x),x)+ \wt{d}_z(x,C)) \notag \\
	    \leq   & ~ 2^{z+2}\cdot d_z(x,c^*(x)) + \frac{2^{2z+2}}{|X_{c^*(x)}|}\cdot (\cost_z(X,C^*) + \cost_z(X,C) ). 
	\end{align}
	where the first step follows from we define $y := \arg\min_{y \in C}d_z(x, y)$, the second step follows from the relaxed triangle inequality (Lemma~\ref{lem:relax_tri_ineq}), the third step follows from $y \in C$ and the definition of $d_z( \wt{c}^*(x),C) )$ (Definition~\ref{def:kz_clustering}), 
	the fourth step follows from Lemma~\ref{lem:approximate_cost}, the fifth step follows from the definition of $X_{c^*(x)}$ (Definition~\ref{def:cluser_set_approx})
	, the sixth step follows from $|X_{\wt{c}^*(x)}| \cdot \wt{d}_z( \wt{c}^*(x),C) ) \le \sum_{y\in X} \wt{d}_z( \wt{c}^*(y),C)$, 
	the seventh step follows from $\wt{c}^*(x)$ is better minimizer than $c^*(x)$ under $\wt{d}$, the eighth step follows from Lemma~\ref{lem:approximate_cost}, and the ninth step follows from the definition of $\wt{\cost}_z$ (Definition~\ref{def:sigma_func_approx}) and $\cost_z$ (Definition~\ref{def:kz_clustering}).

	Thus, we have that
	\begin{align*}
    	    \frac{d_z(x,C)}{\cost_z(X,C)} 
    	\leq & ~ 4 \cdot ( 2^z\cdot \frac{d_z(x,c^*(x))}{\cost_z(X,C)} + \frac{2^{2z}}{|X_{ \wt{c}^*(x)}|}\cdot (1+\frac{\cost_z(X,C^*)}{\cost_z(X,C)}) ) \\
    	\leq & ~ 4 \cdot ( 2^z\cdot \frac{d_z(x,c^*(x))}{\OPT_z} + \frac{2^{2z}}{|X_{ \wt{c}^*(x)}|}\cdot (1+\gamma) ) \\
    	\leq & ~ 4 \cdot ( 2^z\cdot \frac{d_z(x,c^*(x))}{\OPT_z}+ \frac{2^{2z}}{|X_{ \wt{c}^*(x)}|} \cdot 2\gamma ) ,
	\end{align*}
	where the first step follows from Eq.~\eqref{eq:sen1}, the second step follows from $C^*$ is a $(k, \gamma)$-center set (Definition~\ref{def:alhpa_center_set})
	, the third step follows from $\gamma \ge 1$.  
	
	Note that $C$ is arbitrary, thus we complete the proof.
	\end{proof}

\subsection{Coreset Guarantee in Subspace}
\label{sec:correct_output_is_coreset}

In this section, we present the proof on Lemma~\ref{lem:vx12}~that, in a low-dimensional subspace, the output set $D$ of Algorithm~\ref{alg:coreset_gen_alg} is a $\epsilon$-coreset. 
Our lemma gives similar results of Lemma~5.5 in \cite{hv20}, but our setting has more randomness in that we only have approximate metrics on distance and cost function. In this proof, we will show that, the output set $D$ still preserves the cost up to multiplicative $\epsilon$ by the concentration guarantee of our distance queries.

\begin{lemma}[Coreset Guarantee in Subspace]
	\label{lem:vx12}
	Let $\Gamma \in \R^d$ be an arbitrary subspace which is $O(z^{O(z)}\eps^{-z-3} k)$-dimensional.
	We denote $\Gamma'$ to be a subspace which is generated from $\Gamma$ by appending an arbitrary dimension in $\R^d$ which is orthogonal to $\Gamma$. Let $D$ be the set output by Algorithm~\ref{alg:coreset_gen_alg}, and $u$ be the function output by Algorithm~\ref{alg:coreset_gen_alg}.
	Then for any $C\subset \Gamma'$, we have
	\begin{align*}
	     (1 - \eps) \cdot \cost_z(X,C) \le \sum_{x\in D} u(x)\cdot d_z(x,C) \le (1 + \eps) \cdot \cost_z(X,C)
	\end{align*}
	with probability at least $1-\delta/10$.
\end{lemma}

\begin{proof}
	Let $\eps_z$ be the same as in Definition~\ref{def:epsilon_z}. We use $m=O(k/\eps_z)$ to denote the dimension of $\Gamma$. Then, we have,
	\begin{align*}
    	    \sup_{C\in \mathcal{C}} \frac{d_z(x,C)}{\cost_z(X,C)}
    	\leq & ~ 8 \cdot ( 2^z\cdot \frac{d_z(x,c^*(x))}{\OPT_z}+ 2^{2z}\gamma\cdot \frac{1}{|X_{ \wt{c}^*(x)}|} )\\
    	\leq & ~ 8 \cdot ( 2^z\gamma\cdot \frac{d_z(x,c^*(x))}{ \cost_z(X,C^*)}+ 2^{2z} \gamma\cdot \frac{1}{|X_{ \wt{c}^*(x)}|} ) \\
    	\leq & ~ 8 \cdot ( 2^z\gamma\cdot 2 \cdot \frac{ \wt{d}_z(x,c^*(x))}{ \wt{\cost}_z(X,C^*)}+ 2^{2z} \gamma\cdot \frac{1}{|X_{ \wt{c}^*(x)}|} ) \\
    	\leq & ~ \wt{\sigma}(x),
	\end{align*}
	for any $x\in X$, where the first step follows from Claim~\ref{cla:sen_bound}, the second step comes from the definition of $C^*$ (Definition~\ref{def:C_star}), the third step follows from Lemma~\ref{lem:approximate_cost}, 
	the last step comes from the way we define $\wt{\sigma}$ (Definition~\ref{def:sigma_func_approx}). 
	
	Next, we can upper bound the $\sum_{x \in X} \wt{\sigma}(x)$ as follows
	\begin{align*}
	    \sum_{x\in X} \wt{\sigma}(x) 
	= & ~ 2^{2z+4} \cdot \gamma^2\cdot \sum_{x\in X} (\frac{ \wt{d}_z(x,\wt{c}^*(x))}{ \wt{\cost}_z(X,C^*)} + \frac{1}{|X_{ \wt{c}^*(x)}|}) \\
	\leq & ~ 2^{2z+4} \cdot \gamma^2 \cdot (1+k)  \\
	\leq & ~ 2^{2z+6} \cdot \gamma^2 \cdot k,
	\end{align*}
	where the first step follows from the definition of $\wt{\sigma}$ (Definition~\ref{def:sigma_func_approx}) 
	, the second step follows from the definition of $\wt{\cost}_z(X,C^*)$ (Definition~\ref{def:sigma_func_approx}) and $|C^*|=k$, 
	and the last step follows from $k \geq 1$.
	
	Then, we generate $D$ by taking 
	\begin{align*}
	N=\Omega( \epsilon_z^{-2} \sum_{x\in X}\wt{\sigma}(x) \cdot(km \log (\sum_{x\in X}\wt{\sigma}(x))+\log (1/\delta))),
	\end{align*} 
	samples, then applying the second framework of Theorem~\ref{thm:fl11_bfl16} 
	completes the proof.
\end{proof}

\subsection{Approximate Distance}\label{subsec:approx_distance}
In this section, we prove that $\tilde{d}_z$ approximately preserves the distance $d_z$, with multiplicative $(1\pm \epsilon_1)^z$.

\begin{lemma}[Approximate Distance Preserving]\label{lem:approximate_cost}

    Let $\wt{d}_z (x, C)$ denote the approximate distance. 
    Let $d_z(x,C)$ denote the exact distance. 
    Let $\epsilon_1\in (0,0.1)$ denote the error from distance oracle. 
    Then, we have for any $x \in U$ and $C \in {\cal C}_k$
    \begin{align*}
       (1-\epsilon_1)^z \cdot d_z(x,C) \leq \wt{d}_z (x, C) \leq (1+\epsilon_1)^z \cdot d_z(x,C)
    \end{align*}
    and for any $U$ and $C$
    \begin{align*}
        (1-\epsilon_1)^z \cdot \cost_z(U,C) \leq \wt{\cost}_z (U, C) \leq (1+\epsilon_1)^z \cdot \cost_z(U,C).
    \end{align*}
\end{lemma}
Further, choosing $\epsilon_1 = O( \epsilon_{\mathrm{appr}} / z)$ where $\epsilon_{\mathrm{appr}} \in (0,0.01)$ can be chosen to be sufficiently small constant. We obtain $(1\pm\epsilon_{\mathrm{appr}})$-approximation.
\begin{proof}
    This following directly from the definitions of $\wt{d}_z$ (Recall we defined it in Section~\ref{subsec:approx_metrics}) and 
    $\wt{\cost}_z (U, C)$ (Definition~\ref{def:sigma_func_approx}) and approximate distance estimation data structure (Lemma~\ref{lem:ds_dst_est_correct}).
\end{proof}

\subsection{Preservative Property of Subspace}
\label{sec:correct_preserv_prop}

In this section, we present the proof of the preservative property of $(k, z)$-clustering objectives in subspace $\Gamma$. Note that, this lemma is crucial in the proof of the main theorem. This lemma gives similar guarantees to Lemma~5.7 in~\cite{hv20}. Our lemma is different from the previous results in that our setting permits extra randomness on the distance oracle. However, our lemma still holds by Lemma~\ref{lem:approximate_cost} that the distance and cost are still preserved approximately. Hence, the $(k, z)$-clustering objects are still preserved in the subspace $\Gamma$.

\begin{lemma}[Preservative property of subspace $\Gamma$]
	\label{lem:preserve}
	Let $\epsilon_z$ be defined as Definition~\ref{def:epsilon_z}.
	Let $\Gamma \in \R^d$ be a subspace containing $C^*$, i.e., $C^*\subset \Gamma$. For any set $C\in \mathcal{C}_k$ satisfying, 
	\begin{align*}
	    \sum_{x\in X} (d_z(x,\pi(x))-d_z(x,\pi_C(x)))= 0.5 \cdot \epsilon_z \cdot \OPT_z,
	\end{align*}
	where for any $x \in X$, $\pi(x)$ denote the projection of $x$ on $\Gamma$, and $\pi_C(x)$ denote its projection on $\mathrm{Span}(\Gamma\cup C)$.
	Let $D$ be the set output by Algorithm~\ref{alg:coreset_gen_alg}, $u$ be the weight function also output by Algorithm~\ref{alg:coreset_gen_alg}.
	Then we have for any such set $C\in \mathcal{C}$ (Definition~\ref{def:all_center_set})
	\begin{align*}
	        & ~ (1 - 2\eps)\cdot \sum_{x\in D} u(x)\cdot (d_2(\pi(x),C)+d_2(x,\pi(x)))^{z/2}\\
	    \le & ~ \sum_{x\in D} u(x)\cdot d_z(x,C)\\
	    \le & ~ (1 + 2\eps)\cdot \sum_{x\in D} u(x)\cdot (d_2(\pi(x),C)+d_2(x,\pi(x)))^{z/2}
	\end{align*}
	with probability at least $1-\delta/10$.
\end{lemma}

\begin{proof} 
Note that, $\Gamma$ is a subspace of $\R^d$ which contains the set $C^*$, and for any set $C \in {\cal C}$:
	\begin{align*}
	\sum_{x\in X} (d_z(x,\pi(x))-d_z(x,\pi_C(x)))= 0.5 \cdot \epsilon_z \cdot \OPT_z,
	\end{align*}
	
In the first part of the proof, we prove $
	\sum_{x\in D} u(x)\cdot d_z(x,\pi(x)) \leq \sum_{x\in X} d_z(x,\pi(x)) + 0.5 \epsilon_z \cdot \OPT_z$:

To begin with, we have
	\begin{align}\label{eq:lower_bound_d_x_c}
	    \wt{d}_z (x, \wt{c}^*(x)) \geq & ~ 0.5 \cdot d_z (x, \wt{c}^*(x) ) \notag \\
	    \geq & ~ 0.5 \cdot d_z (x , \pi(x) ) 
	\end{align}
	where the first step follows from Lemma~\ref{lem:approximate_cost}, the second step follows from $C^* \subset \Gamma$.
	
	Secondly, we have the following observations:
	\begin{align} \label{eq:upper_bound_cost_X_C} 
	\wt{\cost}_z(X, C^*) 
	\leq & ~ 2  \cost_z(X,C^*) \notag \\ 
	\leq & ~ 2 \gamma\cdot \OPT_z, 
	\end{align}
	where the first step follows from Lemma~\ref{lem:approximate_cost}, the second step follows from  definition of $C^*$ (Definition~\ref{def:alhpa_center_set}).

	Next, we have
	\begin{align}
	\label{eq:sampling2} 
 \wt{\sigma}(x) 
    > & ~ 2^{2z+4} \gamma^2 \cdot \frac{ \wt{d}_z(x, \wt{c}^*(x))}{ \wt{\cost}_z(X,C^*)} \notag \\
    \geq & ~ 2^{2z+4} \gamma^2 \cdot \frac{ 0.5 \cdot d_z(x, \pi(x))}{ \wt{\cost}_z(X,C^*)}  \notag \\
    \geq & ~ 2^{2z+4} \gamma^2 \cdot \frac{ 0.5 \cdot d_z(x, \pi(x))}{ 2 \gamma \OPT_z }  \notag \\
	= & ~ \frac{2^{2z+2} \gamma\cdot d_z(x,\pi(x))}{\OPT_z},  
	\end{align}
	where the first step follows from definition of $\wt{\sigma}(x)$ (Definition~\ref{def:sigma_func_approx}) 
	, the second step follows from Eq.~\eqref{eq:lower_bound_d_x_c}, the third step follows from Eq.~\eqref{eq:upper_bound_cost_X_C}, and the last step follows from merging all the terms.

	Then, we have: 
	\begin{align}
	\label{eq:sampling3}
	\sum_{x\in X} \wt{\sigma}(x) 
	= & ~ 2^{2z+4} \cdot \gamma^2\cdot \sum_{x\in X} (\frac{ \wt{d}_z(x,\wt{c}^*(x))}{ \wt{\cost}_z(X,C^*)}+\frac{1}{|X_{ \wt{c}^*(x)}|}) \notag \\
	\leq & ~ 2^{2z+4} \cdot \gamma^2\cdot(4+k) \notag  \\
	\leq & ~ 2^{2z+6} \cdot \gamma^2 \cdot k. 
	\end{align}
	where the first step follows from the definition of $\wt{\sigma}$ (Definition~\ref{def:sigma_func_approx}), 
	the second step follows from $(|C^*|=k)$, the third step follows from $k \ge 1$ obviously.

	Finally, by Claim~\ref{cla:claim_preserve_1}, we have:
	\begin{align}
	\label{eq:case1}
	    \min_{C\in \mathcal{C}}\sum_{x\in D} u(x)\cdot d_z(x,\pi_C(x))\geq \min_{C\in \mathcal{C}}\sum_{x\in X} d_z(x,\pi_C(x))- 0.5 \cdot \epsilon_z \cdot \OPT_z .
	\end{align}

	In the second part of the proof, we will show that with probability at least $1-\delta/10$, we have the following claim:
	\begin{align}
	\label{eq:case1_2}
	\sum_{x\in D} u(x)\cdot d_z(x,\pi(x)) \leq \sum_{x\in X} d_z(x,\pi(x)) + 0.5 \epsilon_z \cdot \OPT_z .
	\end{align}
	
	At first, we have
	\begin{align}\label{eq:bound_wt_sigma}
	    \frac{\sum_{y\in X} \wt{\sigma}(y) }{ \wt{\sigma}(x) } \leq & ~ \frac{2^{2z+3} \gamma^2k}{{2^{2z+2} \gamma\cdot d_z(x,\pi(x))}/{\OPT_z}} \notag \\
	    = & ~ 2 \gamma k \OPT_z / d_z(x,\pi(x))
	\end{align}
	where the first step follows from 	Eq.~\eqref{eq:sampling2} and Eq.~\eqref{eq:sampling3}, and the final step follows from computation.
	
	Then, note that for each point $x \in D$, we have:
	\begin{align}
	\label{eq:variance}
	|D|\cdot u(x)\cdot d_z(x,\pi(x))
	= & ~ \frac{\sum_{y\in X} \wt{\sigma}(y)}{ \wt{\sigma}(x)}\cdot d_z(x,\pi(x)) \notag \\
	\leq & ~ ( 2 \gamma k \cdot \OPT_z / d_z(x,\pi(x)) ) \cdot d_z(x,\pi(x))  \notag \\
	= & ~ 2\gamma k \cdot \OPT_z. 
	\end{align}
	where first step follows from $|D|\cdot u(x) = \frac{\sum_{y\in X} \wt{\sigma}(y)}{ \wt{\sigma}(x)} $, the second step follows from Eq.~\eqref{eq:bound_wt_sigma}.

	Then by Hoeffding's inequality (Lemma~\ref{lem:hoeffding}) 
	,
	\begin{align*}
    & ~ \Pr \Big[ \big|\sum_{x\in X} d_z(x,\pi(x))-\sum_{x\in D} u(x)\cdot d_z(x,\pi(x)) \big| \geq 0.5 \epsilon_z\cdot \OPT_z \Big] \\
	\leq & ~ 2\cdot \exp \Big( -\frac{2( 0.5 \epsilon_z\cdot \OPT_z )^2}{N \cdot (2\gamma k \cdot \OPT_z)^2} \Big) \\ 
	\leq & ~ \delta/8, 
	\end{align*}
	where the first step follows from Eq.~\eqref{eq:variance}, the second step follows from value of $N$, and the final step follows from calculation.

	Thus, we complete the proof of Eq.~\eqref{eq:case1_2}.

Finally, aggregating the above results gives us the proof of our lemma, i.e., by union bound, with probability at least $1-\delta/10$, Eq.~\eqref{eq:case1} and~\eqref{eq:case1_2}.

	Recall ${\cal C}_k$ is defined as Definition~\ref{def:all_center_set}.
	Then for any $C\in \mathcal{C}_k$,
	\begin{align*}
				& ~\sum_{x\in D} u(x)\cdot d_z(x,\pi(x)) - \sum_{x\in D} u(x)\cdot d_z(x,\pi_C(x)) \\
			\leq & ~ \sum_{x\in D} u(x)\cdot d_z(x,\pi(x)) - \min_{C'\in \mathcal{C}}\sum_{x\in D} u(x)\cdot d_z(x,\pi_{C'}(x)) \\
			\leq & ~ \sum_{x\in X} d_z(x,\pi(x))+ 0.5 \epsilon_z\cdot \OPT_z  -\min_{C'\in \mathcal{C}}\sum_{x\in X} d_z(x,\pi_{C'}(x)) +  0.5 \epsilon_z \cdot \OPT_z  \\
			\leq & ~ \sum_{x\in X} d_z(x,\pi(x)) - \sum_{x\in X} d_z(x,\pi_C(x)) + \epsilon_z\cdot \OPT_z   \\
			\leq & ~ 2\epsilon_z \cdot \OPT_z. 
		\end{align*}
	where the first step follows from taking minimum of the second term, the second step follows from Eq.~\eqref{eq:case1_2}, the third step follows from Eq.~\eqref{eq:case1} and Eq.~\eqref{eq:case1_2}, and the last step follows from assumption.

	In conclusion, applying Lemma~\ref{lem:projection} completes the proof.  
\end{proof}

\subsection{Lower Bound of Sum of Weighted Distance}
\label{sec:weight_dist_close_to_uniform}

Here we present the proof and the statement of Claim~\ref{cla:claim_preserve_1}
used in the proof above. Note that, since we have the \emph{approximate} distance, the function $\wt{\sigma}$ is also an approximation of its exact value. Hence, we need some detailed reasoning about this function. In the following proof, we show that, by the concentration of approximate distances and cost, we can still lower bound the sum of distances produced by the set $D$.

\begin{claim}
	\label{cla:claim_preserve_1}
	Let $\epsilon_z$ be defined as Definition~\ref{def:epsilon_z}. Let $\Gamma \in \R^d$ be a subspace containing $C^*$, i.e., $C^*\subset \Gamma$. For a set $C\in \mathcal{C}$ and for any $x \in X$, let $\pi_C(x)$ be the projection of $x$ on $\mathrm{Span}(\Gamma\cup C)$. We have the following claim,
    	\begin{align}
        	\label{eq:case1_copy}
        	\min_{C\in \mathcal{C}}\sum_{x\in D} u(x)\cdot d_z(x,\pi_C(x))\geq \min_{C\in \mathcal{C}}\sum_{x\in X} d_z(x,\pi_C(x))- 0.5 \cdot \epsilon_z \cdot \OPT_z .
    	\end{align}
	\end{claim}

	\begin{proof}
	
	In this proof, we use $\hat{C} \in {\cal C}_k$ to denote the optimal set minimizing the sum of distances $\sum_{x\in X} d_z(x,\pi_C(x))$, i.e., 
    	\begin{align}
    	\label{eq:def_hat_C_set}
    	    \wh{C} := \arg\min_{C \in {\cal C}_k } \sum_{x\in X} d_z(x,\pi_C(x)).
    	\end{align}
    		
    	To prove the desired inequality, we divide it into two cases.

    	On one hand, if
    	\begin{align*} 
    	\sum_{x\in X} d_z(x,\pi_{\widehat{C}}(x))\leq 0.5  \epsilon_z \cdot \OPT_z ,
    	\end{align*}
    	
    	then, we directly have:
    	\begin{align*}
    	\min_{C\in \mathcal{C}}\sum_{x\in D} u(x)\cdot d_z(x,\pi_C(x))\geq 0\geq \min_{C\in \mathcal{C}}\sum_{x\in X} d_z(x,\pi_C(x))- 0.5  \epsilon_z \cdot \OPT_z .
    	\end{align*}
    	
    	Otherwise, suppose 
    	\begin{align*} 
    	\sum_{x\in X} d_z(x,\pi_{\widehat{C}}(x))> 0.5 \epsilon_z \cdot \OPT_z.
    	\end{align*}
    	
    	Then, due to the optimality of $C^*$ and $C^*\subseteq \Gamma$, for any alternative set $C \subseteq {\cal C}$, we have that:

    	\begin{align*}
    	  \sum_{x\in X} d_z(x,\pi_C(x))\leq \cost_z(X,C^*).
    	\end{align*}
    	
    	Without loss of generality, we can assume $X$ is a point set in the orthogonal complement $\Gamma^\perp$ of the space $\Gamma$. In this light, the span of each set $C \in X$ of size $k$ is a subspace $H \subseteq \Gamma^\perp$, where: 1) the dimension of $H$ is not bigger than $k$. 2) $\mathrm{Span}(\Gamma\cup C) =\mathrm{Span}(\Gamma\cup H)$. It indicates that we can apply Lemma~\ref{lem:subspace} to $\Gamma^\perp$.

    	Recall that $\wt{\sigma}(x)$ is defined as follows (see details in Definition~\ref{def:sigma_func_approx}),
    	\begin{align*}
    	    \wt{\sigma}(x) := 2^{2z+4} \gamma^2 \cdot ( \frac{ \wt{d}_z(x, \wt{c}^*(x))}{ \wt{\cost}_z(X,C^*)} + \frac{1}{ |X_{c^*(x)}| } )
    	\end{align*}
    	
    	Then, by applying Lemma~\ref{lem:subspace} to $\Gamma^\perp$ with setting $\sigma_0$ as:
    	\begin{align*}
    			\sigma_0(x) := & ~ \frac{\wt{\sigma}(x)}{2^{2z+2}\gamma^2} \cdot \frac{ \wt{\cost}_z(X,C^*)}{\sum_{x\in X} d_z(x,\pi_{\widehat{C}}(x))} 
    	\end{align*}
    	We get that, 
    	\begin{align*}
    			\sigma_0(x)  > & ~  \frac{ 4 \wt{d}_z(x,\wt{c}^*(x))}{\sum_{x\in X} d_z(x,\pi_{\widehat{C}}(x))}    \notag \\
    			\geq & ~ \frac{ 2 d(x,\wt{c}^*(x))}{\sum_{x\in X} d_z(x,\pi_{\widehat{C}}(x))}    \notag \\
    			\geq & ~ \frac{ 2 d(x,\pi_C(x))}{\sum_{x\in X} d_z(x,\pi_{\widehat{C}}(x))}    \notag \\
    			\geq & ~ \sup_{C\in \mathcal{C}} \frac{d_z(x,\pi_C(x))}{\sum_{x\in X} d_z(x,\pi_C(x))}. 
    		\end{align*}
    	where the first step follows from Definition of $\wt{\sigma}(x)$ (Definition~\ref{def:sigma_func_approx}), the second step follows from $2 \wt{d}_z(x,\wt{c}^*(x)) \geq {d}_z(x,\wt{c}^*(x))$, the third step follows from $\pi_C$ is a better minimizer than $\wt{c}^*$ under $d$,
    	, and the last step follows from $C^*\in \Gamma$ and Definition of $\widehat{C}$ (see Eq.~\eqref{eq:def_hat_C_set}).

    	We remark that, sampling $x$ from $X$ with probability of $\sigma_0(x)$ is exactly the same as with that of $\wt{\sigma}(x)$. 
    	
    	Additionally, we can show that:
    	\begin{align*}
    			\G 
    			:= & ~\sum_{x\in X} \sigma_0(x) \\
    			= & ~ \sum_{x\in X}  ( \frac{ \wt{d}_z(x, \wt{c}^*(x))}{ \wt{\cost}_z(X,C^*)} + \frac{1}{ |X_{c^*(x)}| } ) \cdot \frac{ \wt{\cost}_z(X,C^*)}{\sum_{x\in X} d_z(x,\pi_{\widehat{C}}(x))} \\
    			= & ~ \sum_{x\in X} \frac{ \wt{d}_z(x, \wt{c}^*(x))}{\sum_{x\in X} d_z(x,\pi_{\widehat{C}}(x))}  +\frac{ \wt{\cost}_z(X,C^*)}{|X_{c^*(x)}|\cdot (\sum_{x\in X} d_z(x,\pi_{\widehat{C}}(x)))} \\
    			= & ~  \frac{ \sum_{x\in X} \wt{d}_z(x,\wt{c}^*(x))}{\sum_{x\in X} d_z(x,\pi_{\widehat{C}}(x))}  +\frac{ k \cdot \wt{\cost}_z(X,C^*)}{   \sum_{x\in X} d_z(x,\pi_{\widehat{C}}(x))} \\
    			\leq & ~ \frac{ 2(k+1)\cdot \cost_z(X,C^*)}{\sum_{x\in X} d_z(x,\pi_{\widehat{C}}(x))}  \\
    			\leq & ~ \frac{ 2 \gamma(k+1)\cdot \OPT_z}{\epsilon_z\cdot \OPT_z/2} \\
    			= & ~ \frac{4 \gamma (k+1)}{\epsilon_z}.
    		\end{align*}
    	where the first step follows from the definition of $\G$, the second step follows from $\sigma_0(x)$ and $\sigma(x)$, the fourth step follows from $|C^*|=k$, the fifth step follows from Lemma~\ref{lem:approximate_cost}, the sixth step follows from $\sum_{x\in X} d_z(x,\pi_{\widehat{C}}(x))> 0.5  \epsilon_z\cdot \OPT_z $, and the last step follows directly.

    	Hence, 
    	\begin{align*} 
    	    N = & ~ O ((  \epsilon_z^{-2} \G^2 )\cdot ( \eps_z^{-1} k^3\log (k/\eps_z)+\log({1}/{\delta}))) \\
    	    = & ~ O( \epsilon_z^{-4} \cdot k^2 \cdot ( \epsilon_z^{-1} k^3 \log(k/\epsilon_z) + \log(1/\delta) ) )
    	\end{align*}
    	as stated in Lemma~\ref{lem:subspace}.

    	By Lemma~\ref{lem:subspace}, we get that the following inequality holds with probability at least $1-\delta/10$,
    	\begin{align*}
    			\min_{C\in \mathcal{C}}\sum_{x\in D} u(x)\cdot d_z(x,\pi_C(x)) 
    		\geq & ~ (1-\frac{\epsilon_z}{2\gamma})\cdot \min_{C\in \mathcal{C}}\sum_{x\in X} d_z(x,\pi_C(x)) \\
    		\geq & ~ \min_{C\in \mathcal{C}}\sum_{x\in X} d_z(x,\pi_C(x))-\frac{\epsilon_z}{2\gamma}\cdot \cost_z(X,C^*) \\
    		\geq & ~ \min_{C\in \mathcal{C}}\sum_{x\in X} d_z(x,\pi_C(x))- 0.5 \epsilon_z\cdot \OPT_z, 
    	\end{align*}
    	where the first step follows from Lemma~\ref{lem:subspace}, the second step follows from $C^* \in \Gamma$, and the last step follows from the guarantee of $C^*$. 
    	
    	Thus, we completes the proof. 
	    
	\end{proof}

\section{Correctness of the Center Set Generation Results} 
\label{sec:center_set_gen_correct}

In this section, we present and prove the correctness of our center set generation algorithm (Algorithm~\ref{alg:center_set_gen}). In Section~\ref{sec:corr_center_gener}, we give the main result and its proof. Since our algorithm has some approximation techniques, we give two different groups of definitions: Section~\ref{sec:center_set_gen_exact_def} states the definitions for exact metrics. Section~\ref{sec:center_set_gen_approx_def} gives the definitions for approximate metrics. Section~\ref{sec:center_gener_cost_bound} shows that the cost of the output map is lower bounded by the output center set. 
We get two outputs from the algorithms: $\tau(x)$ and $V$, $\tau$ maps each $x \in U$ to another point in $U$, this section is saying that, sum of distance between each $x$ and $\tau(x)$ is lower bounded by $\cost_z(U, V)$. 
Section~\ref{sec:center_set_gen_balls_bins} gives the probabilistic discussions of balls and bins, which will be useful for proof. Section~\ref{sec:center_set_gen_covering_set} discusses the size of a special set. Section~\ref{sec:center_set_gen_radius_bound} gives the bound of radius we generated in the algorithm. Section~\ref{sec:center_set_gen_cost_bound} shows that the relation between the cost and the size of the set. Section~\ref{sec:center_set_gen_inte_empt} gives definitions of some subsets we need to discuss and states their intersection is empty. Section~\ref{sec:center_set_gen_size_cost} gives the lower bound of the subset and shows the cost of any subset is lower bounded by its size. Section~\ref{sec:center_set_gen_union_cost} shows that the cost of the union of the subsets is bounded by the sum of the sizes of them. Section~\ref{sec:center_set_gen_size_dec_it} shows that the size of the subset decreased by iteration. Section~\ref{sec:center_set_gen_subset_size_lower} gives the lower bound of the size of the subset. Section~\ref{sec:lower_boud_of_any_center_set} shows that the cost of an arbitrary center set is lower bounded by the parameters we generated in the algorithm.

\subsection{Main Result}
\label{sec:corr_center_gener}

In this section, we present the proof on the correctness of the center set generation results.

\begin{theorem}[Center Set Generation Correctness]
\label{thm:center_set_gen_correct}
    Given a set $U$ of size $n$. Let $\alpha \ge 1$ be a constant. 
    Let $\beta \in (0, 1)$ be a constant. Let $\gamma \in (\beta, 1)$ be a constant. Let $c_0 > 1$ be a constant.  
    The procedure \textsc{CenterSetGen} in Algorithm~\ref{alg:center_set_gen} output an $(k, \Omega(1))$-center set $V$ (Definition~\ref{def:alhpa_center_set}), satisfying
    \begin{itemize}
        \item $|V| = O(k \log(n/k))$;
        \item $\cost_z(U,C) \geq \Omega(1) \cdot \cost_z(U,V)$, for any $C \in \mathcal{C}_k$
    \end{itemize}
    with failure probability at most $e^{-c_0 k}$.

\end{theorem}

\begin{proof}

    We discuss the parameter choices as follows:
    \begin{itemize}
        \item $\alpha \ge 1$ 
        denotes the times we sampling points 
        from the original set.
        The stopping condition of Algorithm~\ref{alg:center_set_gen} also depends on the parameter $\alpha$
        , where we stop the loop in Algorithm~\ref{alg:center_set_gen} if the rest of the set is less than $\alpha k$. To be specific, we choose $\alpha = \Theta(1)$;
        \item $\beta \in (0,1)$ denotes the proportion of a subset taken from the original set when we looking for $v_i$ (for details see Definition~\ref{def:center_set_gen_v_i}). We choose $\beta = \Theta(1)$. 
        \item $\gamma \in (0, 1)$ denotes the parameter we use in definition of $\mu_i$ (for details see Definition~\ref{def:center_set_gen_mu_i}). We choose $\gamma = \Theta(1)$ and $\gamma \in (\beta, 1)$; 
        \item $t \in \mathbb{N}_+$ denotes the times of Line~\ref{line:center_gen_while_loop} in Algorithm~\ref{alg:center_set_gen} loops for. Note that $t = O(\log(n/k))$.
    \end{itemize}
    
    We first note that, for every $i \in [t]$, $V$ will add at most $\lfloor \alpha k \rfloor = O(k)$ points. 
    This implies that 
    \begin{align*}
        |V| = O(t \cdot \lfloor \alpha k \rfloor) = O(k \log(n/k)).
    \end{align*}
    
    We define $r := \lceil\log_{(1-\beta)}((1-\gamma)/3) \rceil$, which is a constant due to we choose $\beta = \Theta(1)$ and $\gamma = \Theta(1)$. 
    
    By choice of $\gamma$ and $r$, we can show
    \begin{align}\label{eq:gamma_r_lower_bound}
        \frac{1-\gamma}{4 r} \geq 0.01
    \end{align}
    
    Then we want to show for any $C \in \mathcal{C}$,
    \begin{align*}
        \Pr[ \cost_z(U, C) \geq \Omega(1) \cdot \cost_z(U, V) ] \geq 1- e^{-c_0 k}.
    \end{align*}
    We can show
    \begin{align*}
         \cost_z(U, C) 
    \ge & ~ \frac{1 - \gamma}{2 r}\sum_{i \in [t]}\mu_i \cdot |C_i| \\
    \ge & ~ \frac{1 - \gamma}{4 r}\sum_{i \in [t]}v_i \cdot |C_i| \\
    \ge & ~ 0.01 \cdot \sum_{i \in [t]}v_i \cdot |C_i| \\
    \ge & ~ 0.01 \cdot \sum_{x \in U}\wt{d}_z(x, \wt{\tau}(x)) \\
    \ge & ~ 0.001 \cdot \sum_{x \in U} d_z(x, \wt{\tau}(x) ) \\
    \ge & ~ 0.001 \cdot \sum_{x \in U} d_z(x, \tau(x) ) \\
    \ge & ~ \Omega(1) \cdot \cost_z(U, V),
    \end{align*}
    where the first step follows from Lemma~\ref{lem:center_set_gen_cost_bound}, the second step follows from $\mu_i \geq v_i/2$ (see Lemma~\ref{lem:center_loop_fail_guarant}), the third step follows from Eq.~\eqref{eq:gamma_r_lower_bound} and the forth step follows from Lemma~\ref{lem:center_set_gen_cost_guarant}, 
    the fifth step follows from $0.5 d \leq \wt{d} \leq 2 d$, the sixth step follows from $\tau$ is minimizer compared to $\wt{\tau}$, the last step follows from Claim~\ref{cla:center_set_gen_cost_guart}.

    Thus we complete the proof.
\end{proof}

 \subsection{Improvement over \texorpdfstring{\cite{mp04}}{}}
 
 \begin{theorem}[\cite{mp04}]
    Given a set $U$ of size $n$. Let $\alpha \ge 1$ be a constant. 
    Let $\beta \in (0, 1)$ be a constant. Let $\gamma \in (\beta, 1)$ be a constant. Let $c_0 > 1$ be a constant.  
    There exists an algorithm that output an $(k, \Omega(1))$-center set $V$ (Definition~\ref{def:alhpa_center_set}), satisfying
    \begin{itemize}
        \item $|V| = O(k \log(n/k))$;
        \item $\cost_z(U,C) \geq \Omega(1) \cdot \cost_z(U,V)$, for any $C \in \mathcal{C}_k$
    \end{itemize}
    with failure probability at most $e^{-c_0 k}$, and runs in time of 
    \begin{align*}
        O(n d k \log (n / k)).
    \end{align*}
 \end{theorem}
 
 We improved the running time in \cite{mp04} to be the following:
 \begin{theorem}[Our result]
    Given a set $U$ of size $n$. Let $\alpha \ge 1$ be a constant. 
    Let $\beta \in (0, 1)$ be a constant. Let $\gamma \in (\beta, 1)$ be a constant. Let $c_0 > 1$ be a constant.  
    There exists an algorithm that output an $(k, \Omega(1))$-center set $V$ (Definition~\ref{def:alhpa_center_set}), satisfying
    \begin{itemize}
        \item $|V| = O(k \log(n/k))$;
        \item $\cost_z(U,C) \geq \Omega(1) \cdot \cost_z(U,V)$, for any $C \in \mathcal{C}_k$
    \end{itemize}
    with failure probability at most $e^{-c_0 k}$, and runs in time of 
    \begin{align*}
        O(z^2 (n k + n d) \log(k) \log(n / k)).
    \end{align*}
 \end{theorem}

\subsection{Exact Definitions}
\label{sec:center_set_gen_exact_def}

Following the same notations as in Algorithm~\ref{alg:center_set_gen}, for proving convenience, we first give the following definitions. We denote by $t$ the times of Line~\ref{line:center_gen_while_loop} in Algorithm~\ref{alg:center_set_gen} loops.

At each iteration, we first define the set of all points as follows.

\begin{definition}[$U_i$]
\label{def:center_set_gen_U_i}
    For any $i \in [t]$, we define the set $U_i$ to be
    \begin{itemize}
        \item $U_0 := U$, the original data set;
        \item $U_i := U_{i-1} \backslash C_{i-1}$, where $C_i$ is defined in Definition~\ref{def:center_set_gen_C_i}, for $i \not= 0$.
    \end{itemize}
\end{definition}

Note in each iteration of Algorithm~\ref{alg:center_set_gen}, we generate a set $S_i$ by sampling points from $U_i$, defined as follows.

\begin{definition}[$S_i$]
\label{def:center_set_gen_S_i}
    For any $i \in [t]$, we define the set $S_i$ to be the set generated by sampling $\lfloor \alpha k \rfloor$ times from $U_i$ (Definition~\ref{def:center_set_gen_U_i}), where $\alpha$ is a constant.
\end{definition}

We then find a real $v_i$ such that it can take specific proportion of the original set $U_i$, defined as follows.

\begin{definition}[$v_i$]
\label{def:center_set_gen_v_i}
    For any $i \in [t]$, let $\beta \in (0, 1)$ be a constant. We define $v_i$ as follows,
    \begin{align*}
        v_i := \min\{ r \in \R ~|~ |B(S_i, r)| \ge \beta \cdot |U_i|\}.
    \end{align*}
\end{definition}

Next we give a name to the set generated from set $S_i$ and radius $v_i$, defined as follows.

\begin{definition}[$C_i$]
\label{def:center_set_gen_C_i}
    Let $S_i$ and $v_i$ are defined as Definition~\ref{def:center_set_gen_S_i} and Definition~\ref{def:center_set_gen_v_i} respectively. For any $i \in [t]$, we define the set $C_i$ as
    \begin{align*}
        C_i := B(S_i, v_i),
    \end{align*}
\end{definition}

Then we generate a map $\tau$, using the distances we have computed before, defined as follows.

\begin{definition}[$\tau$]
\label{def:center_set_gen_tau}
    We define a map $\tau: U\rightarrow U$, where $U$ is the original data set. For every $i \in [t]$, for every $x \in C_i$, we let
    \begin{align*}
        \tau(x) := y,
    \end{align*}
    where $y$ is a point satisfying
    \begin{align*}
        y \in S_i ~\text{and}~ d_z(x, y) \le v_i
    \end{align*}
\end{definition}

The Algorithm~\ref{alg:center_set_gen} finally outputs a set $V$ to be the desired center set. We give its definition here.

\begin{definition}[$V$]
\label{def:center_set_gen_V}
    We define the set $V$ to be,
    \begin{align*}
        V := \bigcup_{i \in [t]}S_i.
    \end{align*}
\end{definition}

In the following proof, we need an important threshold $\mu_i$, here we give its formal definition as follows.

\begin{definition}[$\mu_i$]
\label{def:center_set_gen_mu_i}
    For any $i \in [t]$, let $\mu_i$ denote the minimum non-negative real threshold satisfying that there is a center set $C$ of size 
    $k$ such that the following properties hold:
    \begin{itemize}
        \item the sum of weights for points $x \in U_i$ satisfying $d_z(x,C) \leq \mu_i$ is lower bounded by $\gamma \cdot |U_i|$.
        \item the sum of weights for points $x \in U_i$ satisfying $d_z(x,C) \geq \mu_i$ is lower bounded by $(1-\gamma) \cdot |U_i|$.
    \end{itemize}
    Additionally, such $\mu_i$ is guaranteed to exist.
\end{definition}

\subsection{Approximate Definitions}
\label{sec:center_set_gen_approx_def}

Let $\wt{d}$ denote a fixed metric such that it approximates $d$. We have the following definitions.

At each iteration, we first define the set of all points as follows.

\begin{definition}[$\tilde{U}_i$]
\label{def:center_set_gen_tilde_U_i}
    For any $i \in [t]$, we define the set $\tilde{U}_i$ to be
    \begin{itemize}
        \item $\tilde{U}_0 := U$, the original data set;
        \item $\tilde{U}_i := \tilde{U}_{i-1} \backslash \tilde{C}_{i-1}$, where $\tilde{C}_i$ is defined in Definition~\ref{def:center_set_gen_tilde_C_i}, for $i \not= 0$.
    \end{itemize}
\end{definition}

Note in each iteration of Algorithm~\ref{alg:center_set_gen}, we generate a set $S_i$ by sampling points from $U_i$, defined as follows.

\begin{definition}[$\tilde{S}_i$]
\label{def:center_set_gen_tilde_S_i}
    For any $i \in [t]$, we define the set $\tilde{S}_i$ to be the set generated by sampling $\lfloor \alpha k \rfloor$ times from $\tilde{U}_i$ (Definition~\ref{def:center_set_gen_tilde_U_i}), where $\alpha$ is a constant.
\end{definition}

We then find a real $v_i$ such that it can take specific proportion of the original set $U_i$, defined as follows.

\begin{definition}[$\tilde{v}_i$]
\label{def:center_set_gen_tilde_v_i}
    Let $\beta \in (0, 1)$ be a constant. Then for any $i \in [t]$, 
    We define $\tilde{v}_i$ as follows,
    \begin{align*}
        \tilde{v}_i := \min\{ r \in \R ~|~ |B(\tilde{S}_i, r )| \ge \beta |\tilde{U}_i|\}.
    \end{align*}
\end{definition}

Next we give a name to the set generated from set $S_i$ and radius $v_i$, defined as follows.

\begin{definition}[$\tilde{C}_i$]
\label{def:center_set_gen_tilde_C_i}
    Let $\wt{S}_i$ and $\wt{v}_i$ are defined as Definition~\ref{def:center_set_gen_tilde_S_i} and Definition~\ref{def:center_set_gen_tilde_v_i} respectively.
    For any $i \in [t]$, we define the set $\tilde{C}_i$ as
    \begin{align*}
        \tilde{C}_i := B(\tilde{S}_i, \tilde{v}_i),
    \end{align*}
    where $\tilde{S}_i$ and $\tilde{v}_i$ are defined as Definition~\ref{def:center_set_gen_tilde_S_i} and Definition~\ref{def:center_set_gen_tilde_v_i} respectively. 
\end{definition}

Then we generate a map $\tau$, using the distances we have computed before, defined as follows.

\begin{definition}[$\tilde{\tau}$]
\label{def:center_set_gen_tilde_tau}
    We define a map $\tilde{\tau}: U \rightarrow U$, where $U$ is the original data set. For every $i \in [t]$, for every $x \in \tilde{C}_i$, we let
    \begin{align*}
        \tilde{\tau}(x) := y,
    \end{align*}
    where $y$ is a point satisfying
    \begin{align*}
        y \in \tilde{S}_i ~\text{and}~ \tilde{d}_z(x, y) \le \tilde{v}_i.
    \end{align*}
\end{definition}

The Algorithm~\ref{alg:center_set_gen} finally outputs a set $V$ to be the desired center set. We give its definition here.

\begin{definition}[$\tilde{V}$]
\label{def:center_set_gen_tilde_V}
    We define the set $\tilde{V}$ to be,
    \begin{align*}
        \tilde{V} := \bigcup_{i \in [t]}\tilde{S}_i.
    \end{align*}
\end{definition}

In the following proof, we need an important threshold $\mu_i$, here we give its formal definition as follows.

\begin{definition}[$\tilde{\mu}_i$]
\label{def:center_set_gen_tilde_mu_i}
    For any $i \in [t]$, let $\tilde{\mu}_i$ denote the minimum non-negative real threshold satisfying that there is a center set $C$ of size 
    $k$ such that the following properties hold:
    \begin{itemize}
        \item the sum of weights for points $x \in \tilde{U}_i$ satisfying $\tilde{d}_z(x,C) \leq \tilde{\mu}_i$ is lower bounded by $\gamma \cdot |\tilde{U}_i|$.
        \item the sum of weights for points $x \in \tilde{U}_i$ satisfying $\tilde{d}_z(x,C) \geq \tilde{\mu}_i$ is lower bounded by $(1-\gamma) \cdot |\tilde{U}_i|$.
    \end{itemize}
    Additionally, such $\tilde{\mu}_i$ is guaranteed to exist.
\end{definition}

We first give the following claim about approximation property of the size of set $C$, which will be useful for the proof.

\begin{claim}
\label{cla:approx_claim}
    Let $v_i$, $C_i$, $\wt{v}_i$, $\wt{C}_i$ be defined as Definition~\ref{def:center_set_gen_v_i}, Definition~\ref{def:center_set_gen_C_i}, Definition~\ref{def:center_set_gen_tilde_v_i}, Definition~\ref{def:center_set_gen_tilde_C_i}. We claim
    \begin{align*}
        0.5 \cdot \sum_{i \in [t]}v_i \cdot |C_i| \le \sum_{i \in [t]}\wt{v}_i \cdot |\wt{C}_i| \le 2 \cdot \sum_{i \in [t]}v_i \cdot |C_i|.
    \end{align*}
\end{claim}

\begin{proof}
    Without of generalization, we can assume, fix $i \in [t]$, $S_i = \wt{S_i}$ (Definition~\ref{def:center_set_gen_S_i} and Definition~\ref{def:center_set_gen_tilde_S_i}). Note that we have for any $x \in \wt{U}_i$ and $y \in \wt{S}_i$ we have $0.5 \cdot d_z(z, y) \le \wt{d}_z(x, y) \le 2 \cdot d_z(z, y)$. Recall the way we define $v_i$ and $\wt{v}_i$, we have that
    \begin{align*}
        0.5 \cdot v_i \le \wt{v}_i \le 2 \cdot v_i.
    \end{align*}
    Based on that, recall the way we generate $C_i$ and $\wt{C}_i$, we have that
    \begin{align*}
        0.5 \cdot |C_i| \le |\wt{C}_i| \le 2 \cdot |C_i|.
    \end{align*}
    Thus we have that
    \begin{align*}
        0.5 \cdot \sum_{i \in [t]}v_i \cdot |C_i| \le \sum_{i \in [t]}\wt{v}_i \cdot |\wt{C}_i| \le 2 \cdot \sum_{i \in [t]}v_i \cdot |C_i|.
    \end{align*}
    Thus we complete the proof.
\end{proof}

\subsection{Cost Bounded by the Center Set}
\label{sec:center_gener_cost_bound}

Under the definitions above, now in the rest part of this section, we present some statements for proving the correctness of our center set generation algorithm.

$V$ is the center set output by our Algorithm~\ref{alg:center_set_gen}. We first state the following Claim about its cost bound.

\begin{claim}
\label{cla:center_set_gen_cost_guart}
Let $U$ be the original data set. Let $\tau$ be defined as in Definition~\ref{def:center_set_gen_tau}. Let $V$ be defined as Definition~\ref{def:center_set_gen_V}. Let $d$ and $\cost$ be defined as Definition~\ref{def:kz_clustering}.
    We claim
    \begin{align*}
        \sum_{x \in U} d_z(x, \tau(x) ) \geq \cost_z(U,V).
    \end{align*}
\end{claim}

\begin{proof}
    Notice that, by the way we generate $\tau$ (Definition~\ref{def:center_set_gen_tilde_tau}), we have for any $x \in U$, $\tau(x) \in S_i$ for some $i \in [n]$. 
    
    Then, by the way we define $V$ (Definition~\ref{def:center_set_gen_V}), we know $S_i \subset V$, $\forall i$. 
    
    Thus we have for any $x \in U$, $\tau(x) \in V$.
    
    Then we have
    \begin{align*}
         \sum_{x \in U} d_z(x, \tau(x) ) 
    \ge & ~ \sum_{x \in U} d_z(x, U) \\
    =   & ~ \cost_z(U,V),
    \end{align*}
    where the first step follows from $\tau(x) \in V$, the second step follows from the defintion of $\cost_z(U, V)$.
    
    Thus we complete the proof.
\end{proof}
    
\subsection{Balls and Bins Discussion}
\label{sec:center_set_gen_balls_bins}

Here in this section, we introduce some probability discussion will be used in the following proofs. 

Keeping that for now, we first define two functions as follows.

\begin{definition}[Function $f(m, a, b)$]
\label{def:center_set_gen_func_f}
    For any positive integer $m$ and any non-negative reals $a$ and $b$, we define $f(m, a, b)$ to be the probability that more than $a m$ bins remain empty after $\lceil b \rceil$ balls are thrown at random (uniformly and independently) into $m$ bins. 
\end{definition}    

\begin{definition}[Function $g(m, a, b, v)$]
\label{def:center_set_gen_func_g}
    For any positive integer $m$, non-negative reals $a$ and $b$, and $m$-vector $v = (r_0, \dots, r_{m-1})$ of non-negative reals $r_i$, we define $g(m, a, b, v)$ as follows. Consider a set of $m$ bins numbered from $0$ to $m - 1$ where bin $i$ has associated weighted $r_i$. Let $R$ denote the total weight of the bins. Assume that each of $\lceil b \rceil$ balls is thrown independently at random into one of the $m$ bins, where bin $i$ is chosen with probability $r_i/R$, $0 \le i < m$. We define $g(m, a, b, v)$ as the probability that the total weight of the empty bins after all balls have been thrown is more than $a R$.
\end{definition}

Then we introduce the following lemma, giving the upper bound of function $g$.

\begin{lemma}[Lemma~3.2 in \cite{mp04}]
\label{lem:center_set_gen_intro_lemma}
    Let $g(m, a, b, v)$ be defined as Definition~\ref{def:center_set_gen_func_g}. For any positive real $\epsilon$ 
    , there is a positive real $\lambda$ such that for all positive integers $m$ and any real $b \ge m$, we have $g(m, \epsilon, \lambda b, v) \le e^{-b}$  for all $m$-vectors $v$ of non-negative reals.
\end{lemma}

\subsection{Covering Set Size Guarantee} 
\label{sec:center_set_gen_covering_set}

Expect from the probabilistic discussion, we still need some other tools. Here we define a specific relationship called ``covering'', and we claim the covering set has as size lower bound. This claim is crucial to the proof of Lemma~\ref{lem:center_loop_fail_guarant}.

\begin{claim}
\label{cla:center_set_gen_fail_cla}
    Let $U_i$ be defined as Definition~\ref{def:center_set_gen_U_i}. Let $\mu_i$ be defined as Definition~\ref{def:center_set_gen_mu_i}. Let $S_i$ be defined as Definition~\ref{def:center_set_gen_S_i}. Let $C$ be a set of $k$ points such that $|B(X, \mu_i)| \ge \gamma |U_i|$. We define the set
    \begin{align*}
        G :=  U_i \cap B(X, \mu_i).
    \end{align*}
    For each $c \in C$, we define $X_{c}$ to be the set
    \begin{align*}
        X_c := \{y \in G ~|~ \arg\min_{c \in C}d_z(y, c) \}.
    \end{align*}
    For any $c \in C$, we say that $S_i$ covers $X_c$ iff $S_i \cap X_c$ is nonempty. Let $G'$ denote the set of points covered by $S_i$. Note that $G' \subseteq G$.
    
    Then we have that, for any positive reals $\epsilon$ and $c_0$, there is a sufficiently large choice of $\alpha$ such that $|G'| \ge (1 - \epsilon)|G|$ with probability at least $1 - e^{- c_0 k}$. 
\end{claim}

\begin{proof}
    First, by the definition of $\mu_i$ (Definition~\ref{def:center_set_gen_mu_i}) we can deduce that at least a $\gamma$ fraction of the total set is associated with points in $G$. Thus, by Lemma~\ref{lem:chernoff_bound} we have that for any positive reals $\lambda$ and $c_0$, there exists a sufficiently large choice of $\alpha$ such that at least $\lambda k$ of the $\lfloor \alpha k \rfloor$ samples associated with the construction of $S_i$ are good with probability of failure at most $e^{- c_0 k} / 2$.
    
    To ensure the $|G'|$ is at least $(1 - \epsilon)|G|$ with failure probability at most $e^{- c_0 k} / 2$, we can apply Lemma~\ref{lem:center_set_gen_intro_lemma} by viewing each sample associated with a point in $G \cap S_i$ as a ball toss and each set $X_c$ as a bin with uniform weight. The claim then follows.
\end{proof}

\subsection{The Radius is Upper Bounded}
\label{sec:center_set_gen_radius_bound}

The following lemma established the main probabilistic claim used in our analysis of Algorithm~\ref{alg:center_set_gen}. We note that the lemma holds with high probability by taking $\alpha$ and $\beta$ appropriately.

\begin{lemma}[Lemma~3.3 in \cite{mp04}]
\label{lem:center_loop_fail_guarant}
    Let $\mu_i$ and $v_i$ be defined in Definition~\ref{def:center_set_gen_mu_i} and Definition~\ref{def:center_set_gen_v_i} respectively. Let $c_0$ be any positive real. Let $\alpha > 1$ be a constant. 
    There exists a sufficiently large choice of $\alpha$ such that $v_i \le 2 \mu_i$ for all $i \in [t]$  
    , with failure probability at most $e^{-c_0 k}$. 
\end{lemma}

For the completeness of our paper, we still provide a proof.
\begin{proof}
    We notice that $\beta$ (the factor appearing in the definition of $v_i$, see Definition~\ref{def:center_set_gen_v_i}) is less than $\gamma$ (remember we set $\gamma \in (\beta, 1)$), and for all points $y$ covered by $S_i$ we have $d_z(y, S_i) \le 2 \mu_i$.
    
    Then by Claim~\ref{cla:center_set_gen_fail_cla} we have the Lemma proved.
\end{proof}

\subsection{Bounding the Cost}
\label{sec:center_set_gen_cost_bound}

In this section, we gives an upper bound of the cost of the map $\tau$ we generated in Algorithm~\ref{alg:center_set_gen}. 

\begin{lemma}[Modified from Lemma~3.5 in \cite{mp04}]
\label{lem:center_set_gen_cost_guarant}
    Let $\tau$, $v_i$, and $C_i$ be defined as Definition~\ref{def:center_set_gen_tau}
    , Definition~\ref{def:center_set_gen_v_i} and Definition~\ref{def:center_set_gen_C_i} respectively. We have 
    \begin{align*}
        \sum_{x \in U}\wt{d}_z(x, \wt{\tau}(x)) \le 2 \cdot \sum_{i \in [t]}v_i \cdot |C_i|,
    \end{align*}
    where the $t$ denotes how many times it loops.
\end{lemma}

\begin{proof}
    For any $i \in [t]$, observe that
    \begin{align*}
         \sum_{x \in U}\wt{d}_z(x, \wt{\tau}(x))
    =   & ~ \sum_{i \in [t]}\sum_{x \in \wt{C}_i} \wt{d}_z(x, \wt{\tau}(x)) \\
    \le & ~ \sum_{i \in [t]}\sum_{x \in\wt{C}_i} \wt{v}_i\\
    =   & ~ \sum_{i \in [t]}\wt{v}_i \cdot |\wt{C}_i| \\
    \le & ~ 2 \cdot \sum_{i \in [t]}v_i \cdot |C_i|,
    \end{align*}
    where the first step follows from that we cut each $\wt{C}_i$ from $U$, and the union of $\wt{C}_i$ over $i$ is $U$, the second step follows from the way we construct $\wt{\tau}(x)$ 
    , the third step follows immediately, and the last step follows from Claim~\ref{cla:approx_claim}. 
    
    Thus we complete the proof.
\end{proof}

\subsection{Intersection of each Subset is Empty}
\label{sec:center_set_gen_inte_empt}

In this section, we focus on establishing a lower bound on the cost of any center set. Throughout the remainder of this section we fix an arbitrary center set $C$ of size $k$. We define some sets here.

\begin{definition}
\label{def:center_set_gen_sets}
    For any $i \in [t]$, we define the set
    \begin{align*}
        F_i := \{x \in U_i ~|~ d_z(x, C) \ge \mu_i\}.
    \end{align*}
    For any positive integer $m$, let $u := \lfloor \frac{t - i}{j} \rfloor$. Let $F_i^m$ denote the set $F_i\backslash(\cup_{j \in [u]}F_{i + j m})$ and we let $G_{i, m}$ denote the set of all integers $j$ such that $0 \le j \le t$ and $j \equiv i (\mathrm{mod}~m)$.
\end{definition}
 We still use $t$ to denote the time of the algorithm loops here in this section. The following lemma claims that the $F$ set of each iteration is empty.

\begin{lemma}[Lemma~3.6 in \cite{mp04}]
\label{lem:F_set_discuss}
    Let $i$, $j$, $l$ and $m$ be integers such that $0 \le l \le t$, $m > 0$, $i \not= j$ and $i,j \in G_{l,m}$. Then $F_i^m \cap F_j^m = \emptyset$.
\end{lemma}

\subsection{Set Size and Cost Guarantee}
\label{sec:center_set_gen_size_cost}

We here present the following lemma, showing that for arbitrary center set $C$, the cost of any subset of each $F_i$ is lower bounded by its size, also saying the size of $F_i$ is lower bounded by size of $U_i$. 

\begin{lemma}[Lemma~3.7 in \cite{mp04}]
\label{lem:cost_bound_by_subset}
    Let $U_i$ and $\mu_i$ be defined as Definition~\ref{def:center_set_gen_U_i} and Definition~\ref{def:center_set_gen_mu_i}. For integer $i \in [t]$, let $Y$ be a subset of $F_i$. Then $|F_i| \ge (1 - \gamma) \cdot |U_i|$ and $\cost_z(Y, C) \ge \mu_i \cdot |Y|$. 
\end{lemma}

\begin{proof}
    We first note that by the definition of $\mu_i$ (Definition~\ref{def:center_set_gen_mu_i}), $|F_i|$ is at least $(1 - \gamma) \cdot |U_i|$. By the definition of $F_i$, we have that $d_z(y, C) \ge \mu_i$ for any $y \in F_i$. Thus we have
    \begin{align*}
        \cost_z(Y, C) = \sum_{y \in Y}d_z(y, C) \ge \mu_i \cdot |Y|.
    \end{align*}
    Thus we complete the proof.
\end{proof}

\subsection{Cost of Union Set is Bounded by Set Sizes}
\label{sec:center_set_gen_union_cost}

Here in this section, we are going to show that, the cost of the union of the subsets $F_i^m$ is lower bounded by the sum of their sizes multiplied by the special threshold $\mu_i$ (Definition~\ref{def:center_set_gen_mu_i}).

\begin{lemma}[Lemma~3.8 in \cite{mp04}]
\label{lem:cost_to_F_set}
    Let $F$ and $G$ defined as Definition~\ref{def:center_set_gen_sets}. Let $\mu_i$ be defined as Definition~\ref{def:center_set_gen_mu_i}. For all integers $0 \le l \le t$ and $m > 0$, we have
    \begin{align*}
        \cost_z(\bigcup_{i \in G_{l,m}}F_i^m, C) \ge \sum_{i \in G_{l,m}} \mu_i \cdot |F_i^m|.
    \end{align*}
\end{lemma}

\begin{proof}
    By Lemma~\ref{lem:F_set_discuss}, for all $l$ and $m$ such that $0 \le l \le t$ and $m > 0$, we have 
    \begin{align*}
         \cost_z(\bigcup_{i \in G_{l,m}}F_i^m, C) = \sum_{i \in G_{l,m}}\cost_z(F_i^m, C).
    \end{align*}
    Then by Lemma~\ref{lem:cost_bound_by_subset} we have
    \begin{align*}
        \cost_z(F_i^m, C) \ge \mu_i \cdot |F_i^m|.
    \end{align*}
    Thus the claim follows.
\end{proof}

\subsection{Set Size is Decreased by Iteration}
\label{sec:center_set_gen_size_dec_it}

Let $\beta$ and $\gamma$ be defined in Theorem~\ref{thm:center_set_gen_correct}. For the remainder of this section, let $r:=\lceil \log_{1-\beta}((1 - \gamma)/3)\rceil$.

\begin{lemma}[Lemma~3.9 in \cite{mp04}]
\label{lem:center_set_gen_set_size}
    For all $i$ such that $0 \le i \le t$, we have $|F_{i+r}| \le \frac{1}{3} \cdot |F_i|$.
\end{lemma}

\begin{proof}
    Note that
    \begin{align*}
            |F_{i + r}|
    \le & ~ |U_{i + r}|\\
    \le & ~ (1 - \beta)^r \cdot |U_i|\\
    \le & ~ \frac{(1 - \beta)^r}{1 - \gamma} \cdot |F_i|,
    \end{align*}
    where the first step follows from $U_{i+r} \subseteq F_{i+r}$, the second step follows from the way we generate $U_i$ (Definition~\ref{def:center_set_gen_U_i}), the last step follows from Lemma~\ref{lem:cost_bound_by_subset}.
    
    Note $\frac{(1 - \beta)^r}{1 - \gamma} = \frac{1}{3}$, thus we complete the proof.
\end{proof}

\subsection{Subset Size Lower Bound}
\label{sec:center_set_gen_subset_size_lower}

Here in this section, we present the following lemma, by setting $r$ as $\lceil \log_{1-\beta}((1 - \gamma)/3)\rceil$, we have the the size of $F_i^r$ is lower bounded by a half of the size of set $F_i$.

\begin{lemma}[Lemma~3.10 in \cite{mp04}]
\label{lem:center_set_gen_set_prop}
    Let $F$ be defined as Definition~\ref{def:center_set_gen_sets}. For all $i \in [t]$, we have 
    \begin{align*}
        |F_i^r| \ge \frac{|F_i|}{2}.
    \end{align*}
\end{lemma}

\begin{proof}

    Let $u := \lfloor \frac{t - i}{r} \rfloor$. Observe that
    \begin{align*}
            |F_i^r|
    =   & ~ |F_i \backslash \cup_{j \in [u]}F_{i + j r}| \\
    \ge & ~ |F_i| - \sum_{j \in [u]}\frac{|F_j|}{3^j} \\
    \ge & ~ \frac{|F_i|}{2},
    \end{align*}
    where the first step follows from the definition of $F_i^r$ (Definition~\ref{def:center_set_gen_sets}), the second step follows from Lemma~\ref{lem:center_set_gen_set_size}, and the last step follows from $\sum_{j = 1}^{n} \frac{1}{3^j} = \frac{1}{2}$.
    
    Thus we complete the proof.
\end{proof}

\subsection{Cost Lower Bound for Arbitrary Center Set}
\label{sec:lower_boud_of_any_center_set}

In this section, we present the following lemma, which gives the the lower bound of cost for arbitrary center set.

\begin{lemma}[Lemma~3.11 in \cite{mp04}]
\label{lem:center_set_gen_cost_bound}
    Let $\mu_i$ and $C_i$ be defined as Definition~\ref{def:center_set_gen_mu_i} and Definition~\ref{def:center_set_gen_C_i} respectively.
    We define 
    \begin{align*}
        r:=\lceil \log_{(1-\beta)}((1-\gamma)/3) \rceil.
    \end{align*}
    Recall that $t$ is the times of the algorithm loops.
    Then we have for any set $C \in \mathcal{C}$ that
    \begin{align*}
        \cost_z(U, C) \ge \frac{1 - \gamma}{2r} \cdot \sum_{i \in [t]}\mu_i \cdot |C_i|,
    \end{align*}
    
\end{lemma}

\begin{proof}
    Let $l := \arg\max_{0 \le j < r}\{\sum_{i \in G_{j, r}}|F_i^r|\}$ and fix center set $C$ of size $k$. The cost $\cost_z(U, C)$ is at least
    \begin{align*}
            \cost_z(\bigcup_{i \in G_{l,r}}F_i^r, C)
    \ge & ~ \sum_{i \in G_{l,r}}\mu_i|F_i^r| \\
    \ge & ~ \frac{1}{r} \sum_{i \in [t]}\mu_i|F_i^r| \\
    \ge & ~ \frac{1}{2 r} \sum_{i \in [t]}\mu_i|F_i| \\
    \ge & ~ \frac{1 - \gamma}{2 r} \sum_{i \in [t]}\mu_i|U_i| \\
    \ge & ~ \frac{1 - \gamma}{2 r} \sum_{i \in [t]}\mu_i|C_i|,
    \end{align*}
    where the first step follows from Lemma~\ref{lem:cost_to_F_set}, the second step follows from averaging and the choice of $l$, the third step follows from Lemma~\ref{lem:center_set_gen_set_prop}, the fourth step follows from Lemma~\ref{lem:cost_bound_by_subset}, and the last step follows from $C_i \subseteq U_i$.
    
    Thus we complete the proof.
\end{proof}

\section{Improvement over \texorpdfstring{\cite{css21}}{}}
\label{sec:improve_cass21}

In \cite{css21}, the authors present a new coreset generation algorithm. This framework simultaneously improves upon the coreset size bound for some different settings. To compared with the result in \cite{hv20}, let $\mathcal{K} := \min\{\epsilon^{-2} + \epsilon^{-z}, k\epsilon^{-2}\} \cdot \poly\log(\epsilon^{-1}) $. The framework presented in \cite{css21} achieves a coreset size of $O(\mathcal{K}\cdot k \cdot \min\{d, \epsilon^{-2} \log k\})$ in $d$-dimensional Euclidean space, better than the coreset size of $O(k \log k \cdot \epsilon^{-2z} \cdot \min\{d, \epsilon^{-2} \log k\})$, improving the $\log k$ factor.

The main technique in \cite{css21} is the \emph{structured partition}. By dividing the original dataset into different \emph{groups}, the algorithm can significantly improve the coreset size. 

The algorithm is constructing a coreset based on a center set preserving the cost. So the running time is dominated by the center set generation time, which is $O(n d k)$ in \cite{css21} because they use original center set generation algorithm presented in \cite{mp04}. This leads the running time of their algorithm to $\wt{O}(n d k)$. After getting the desired center set, the algorithm runs the partitioning procedure and the following steps. In their design of the algorithm, the following operations takes $\wt{O}(n)$ time. Recall that we have improved the running time of the center set generation algorithm to $\wt{O}(n k + n d)$, which implies that, we can plug this algorithm into the coreset generation algorithm. After combining the algorithms, we have the running time improved to $\wt{O}(n d + n k)$. 

We first give some preliminaries for our algorithm in Section~\ref{sec:cass21_prel}, including definitions and some useful lemmas from previous work. Then in Section~\ref{sec:cass21_alg} we present the algorithm formally, and give the proof for its correctness and running time. 

\subsection{Preliminary}
\label{sec:cass21_prel}

Here we present some preliminaries for the algorithm design.  

\subsubsection{Partitioning Structure: Groups}

As stated before, the algorithm partitions the original dataset into different groups. We present related definitions here.

\begin{definition}
\label{def:avrg_cost}
    For any $i \in [k]$, we denote the average cost of a cluster $\mathbb{C}_i$ as
        \begin{align*}
            \Sigma_{\mathbb{C}_i} := \frac{\cost_z(\mathbb{C}_i, C)}{|\mathbb{C}_i|}.
        \end{align*}
\end{definition}

\begin{definition}[Ring definitions, follows the definitions in \cite{css21}]
\label{def:ring_defs}
    Fix a center set $C$ of size $k$, let $\mathbb{C}_1, \dots, \mathbb{C}_k$ be the clusters induces by $C$. Let $\Sigma$ be defined as Definition~\ref{def:avrg_cost}. We define the following:
    \begin{itemize}
        \item For any $i \in [k]$ and $j \in \Z$, we define the \emph{ring} $R_{i,j}$ to be a set that
        \begin{align*}
            R_{i,j} := \{x \in \mathbb{C}_i ~|~ 2^j \cdot \Sigma_{\mathbb{C}_i} \le \cost_z(x, C) \le 2^{j + 1} \cdot \Sigma_{\mathbb{C}_i}\}.
        \end{align*}
        
        \item For any $i \in [k]$, we define the \emph{inner ring} of a cluster $\mathbb{C}_i$ to be
        \begin{align*}
            R_I(\mathbb{C}_i) := \bigcup_{j \le 2z\log(\epsilon/z)\mathrm{~and~}j\in\Z}R_{i, j}.
        \end{align*}
        Similarly. we define the \emph{outer ring} of a cluster $\mathbb{C}_i$ to be
        \begin{align*}
            R_O(\mathbb{C}_i) := \bigcup_{j > 2z\log(\epsilon/z)\mathrm{~and~}j\in\Z}R_{i, j}.
        \end{align*}
        We also define the \emph{main ring} to be the rest of points in $\mathbb{C}_i$:
        \begin{align*}
            R_M(\mathbb{C}_i) := \mathbb{C}_i \backslash R_I(\mathbb{C}_i) \backslash R_O(\mathbb{C}_i).
        \end{align*}
        For a center set $C$ of size $k$, we define
        \begin{align*}
            R_I^C := \bigcup_{i \in [k]} R_I(\mathbb{C}_i)\\
            R_O^C := \bigcup_{i \in [k]} R_O(\mathbb{C}_i)
        \end{align*}
        
        \item For each $j \in \Z$, we define 
        \begin{align*}
            R_j := \bigcup_{i \in [k]}R_{i, j}.
        \end{align*}
        
        \item For each $j \in \Z$ and $b \in \Z$, we define the \emph{group} as
        \begin{align*}
            G_{j, b} := \{x | \exists i, x \in R_{i, j} \mathrm{~and~} (\frac{\epsilon}{4z})^z \cdot \frac{\cost_z(R_j, C)}{k} \cdot 2^b \le \cost_z(R_{i,j}, C) \le (\frac{\epsilon}{4z})^z \cdot \frac{\cost_z(R_j, C)}{k} \cdot 2^{b + 1}\}.
        \end{align*}

    \end{itemize}
\end{definition}

Following the definition of Rings, we introduce the definition of groups as follows.

\begin{definition}[Group definitions, follows the definitions in \cite{css21}]
\label{def:group_defs}
    Let $R$ be defined as Definition~\ref{def:ring_defs}. We have the following definitions:
    \begin{itemize}
        \item For any $j \in \Z$, define the union of the cheapest groups as
        \begin{align*}
            G_{j, \min} := \bigcup_{b \le 0}G_{j, b}.
        \end{align*}
        Similarly, we define the union of the most expensive groups as
        \begin{align*}
            G_{j, \max} := \bigcup_{b \ge z \log(\frac{4z}{\epsilon})}G_{j, b}.
        \end{align*}
        We say the set of $G_{j, \min}$, $G_{j, \max}$ and all $G_{j,b}$ that $b \in (0, z \log(4z/\epsilon))$ is the \emph{interesting groups}.
        
        \item Let $b \in \Z$. The set of outer rings is also partitioned into \emph{outer groups}:
        \begin{align*}
            G_b^O := \{x | \exists i, x \in \mathbb{C}_i \mathrm{~and~} (\frac{\epsilon}{4z})^z \cdot \frac{\cost_z(R_O^C, C)}{k} \cdot 2^b \le \cost_z(R_{i,j}, C) \le (\frac{\epsilon}{4z})^z \cdot \frac{\cost_z(R_O^C, C)}{k} \cdot 2^{b + 1}\}.
        \end{align*}
        
        \item Similarly, we define as well
        \begin{align*}
            G_{\min}^O := \bigcup_{b < 0 \mathrm{~and~} b \in \Z}G_b^O
        \end{align*}
        and
        \begin{align*}
            G_{\max}^O := \bigcup_{b \ge z \log(\frac{4z}{\epsilon}) \mathrm{~and~} b \in \Z}G_b^O.
        \end{align*}
    \end{itemize}
    
\end{definition}

We here list some fact about the partitioning.

\begin{fact}[Fact~1 in \cite{css21}]
    There are at most $O(z \log (z/\epsilon))$ non-empty $R_j$'s that are not in some $R_I^C$ or $R_O^C$.
\end{fact}

Thus we have that, the non-empty groups count is also bounded:

\begin{fact}[Fact~2 in \cite{css21}]
    There are at most $O(z^z \log^2(z/\epsilon))$ interesting groups $G_{j,b}$.
\end{fact}

From the definition of outer groups, we have that 
\begin{fact}[Fact~3 in \cite{css21}]
    There are at most $O(z \log (z/\epsilon))$ interesting outer groups.
\end{fact}

\subsubsection{Useful Lemmas from \texorpdfstring{\cite{css21}}{}}

Here we present some prior results from \cite{css21}, which will be useful for our coreset generation algorithm. 

We first introduce an approximation relation between two center sets.

\begin{definition}[Definition~1 in \cite{css21}]
\label{def:approx_center_set}
    Let $X \subseteq \R^d$ be the original dataset of $n$ points. Let $C_1 \subseteq \R^d$ be a center set. Let $k$, $z$ be positive integers. We say a center set $C_2 \subseteq \R^d$ is a $C_1$-approximate center set for $(k, z)$-clustering if the following statement is satisfied:
    
    For every set $C_0 \subseteq \R^d$ of size $k$, there exists a set $\wt{C}_0 \subseteq \C_2$ of size $k$ such that for all $x \in X$ that satisfies 
    \begin{align*}
        \cost_z(x, C_0) \le (\frac{8z}{\epsilon})^z \cdot \cost_z(x, C_1) \mathrm{~or~} \cost_z(x, \wt{C}_0) \le (\frac{8z}{\epsilon})^z \cdot \cost_z(x, C_1),
    \end{align*}
    the following holds
    \begin{align*}
        |\cost_z(x, C_0) - \cost_z(x, \wt{C}_0)| \le \frac{\epsilon}{z \log(z/\epsilon)} \cdot (\cost_z(x, C_0) + \cost_z(x, C_1)).
    \end{align*}
\end{definition}

Next, We present the following lemma introducing the \textsc{GroupSample} procedure from \cite{css21}.

\begin{lemma}[\textsc{GroupSample} algorithm, Lemma~2 in \cite{css21}]
\label{lem:group_sample}
    Let $k$, $z$ be positive integers. Let $G$ be a group of points (Definition~\ref{def:group_defs}) and $C_0 \subset \R^d$ be a center set of size $k$ such that,
    \begin{itemize}
        \item For every cluster $\mathbb{C}$ induced by $C_0$ on $G$, we have that
        \begin{align*}
            \forall x, y \in \mathbb{C},~\cost_z(x, C_0) \le 2 \cost_z(y, C_0).
        \end{align*}
        
        \item For all cluster $\mathbb{C}$ induced by $C_0$ on $G$, we have that
        \begin{align*}
            \cost_z(\mathbb{C}, C_0) \ge \frac{\cost_z(G, C_0)}{2k}.
        \end{align*}
    \end{itemize}
    Let $C_1$ be a $C_0$-approximate center set for $(k,z)$-clustering on $G$ (Definition~\ref{def:approx_center_set}). Let $\beta > 0$ be a positive integer. Then there is an algorithm named \textsc{GroupSample}, takes $G$ and $\beta$ as input, and constructs a set $D$  of size $\beta$ such that for all set $C \in \R^d$ of size $k$, the following holds
    \begin{align*}
        |\cost_z(G, C) - \cost_z(D, C)| = O(\epsilon) \cdot (\cost_z(G, C) + \cost_z(G, C_0)),
    \end{align*}
    with probability 
    \begin{align*}
        1 - \exp(k \log|C_1| - 2^{O(z \log z)} \cdot \frac{\min\{\epsilon^2, \epsilon^z\}}{\log^2(1/\epsilon)} \cdot \beta),
    \end{align*}
    and runs in time 
    \begin{align*}
        O(|G| \cdot d).
    \end{align*}
\end{lemma}

We also need a lemma describing another subroutine of the algorithm.

\begin{lemma}[\textsc{SensitivitySample} algorithm, Lemma~3 in \cite{css21}]
\label{lem:senstivity_sample}
    Let $X \subseteq \R^d$ be a dataset of $n$ points. Let $k$, $z$ be positive integers. Let $C_0 \subset \R^d$ be a $(k, \alpha)$-center set of size $k$ (Definition~\ref{def:alhpa_center_set}). 
    
    Let $G$ be either a $G_b^O$ or a $G_{\max}^O$. Suppose $C_1$ to be a $C_0$ approximate center set on $G$ (Definition~\ref{def:approx_center_set}). 
    
    Let $\beta$ be a positive integer. Then there is an algorithm \textsc{SensitivitySample}, takes $G$ and $\beta$ as input, and constructs a set $D$ of size $\beta$ such that for all set $C \in \R^d$ of size $k$, the following holds,
    \begin{align*}
        |\cost_z(G, C) - \cost_z(D, C)| = \frac{z}{z \log(z/\epsilon)} \cdot (\cost_z(X, C) + \cost_z(X, C_0)),
    \end{align*}
    with probability 
    \begin{align*}
        1 - \exp(k \log |C_1| - 2^{O(z \log z)} \cdot \frac{\epsilon^2}{\log^2(1 /\epsilon)} \cdot \beta),
    \end{align*}
    and runs in time
    \begin{align*}
        O(|G| \cdot d).
    \end{align*}
\end{lemma}

\subsection{An Improved Coreset Generation Algorithm over \texorpdfstring{\cite{css21}}{}}
\label{sec:cass21_alg}

Here we state the formal coreset generation algorithm by plugging our center set generation algorithm to the result of \cite{css21}. We then state the correctness of the algorithm, together with the running time analysis here.

\begin{algorithm}[!ht]\caption{Modified Coreset Generation from \cite{css21}}
\label{alg:coreset_gen_css21}
    \begin{algorithmic}[1]
    \Procedure{CoresetGen}{$X, n, d, \epsilon, \delta \in (0, 0.1), z \ge 1, k \ge 1$} \Comment{Theorem~\ref{thm:main_correct}, and ~\ref{thm:main_runtime}}
        \State $\gamma \gets O(1)$
        \State $C^* \gets \textsc{CenterSetGen}(\gamma,\delta, n, d, U)$\label{line:css_coreset_center_set_gen}\Comment{Algorithm~\ref{alg:center_set_gen}}
        
        \State Initiate a map $\sigma: X \rightarrow \R$ \Comment{The weight function of the coreset}
        
        \For{$c \in C^*$} \label{line:css_coreset_sigma_reset}
            \State $\sigma(c) \gets 0$
        \EndFor
        \State $P_0, P_1 \gets \emptyset$ \Comment{Used for discarded points and positive weighted points respectively}
        
        \For{$c \in C^*$}\label{line:css_coreset_avrg_cost}
            \State Compute $\Sigma_{\mathbb{C}_c}$ 
            \Comment{Definition~\ref{def:avrg_cost}}
        \EndFor
        \State Partition $X$ into groups $R_j$\label{line:css_coreset_partition} \Comment{Definition~\ref{def:ring_defs}, using the $\Sigma_{\mathbb{C}_c}$'s computed last step}
        
        \For{$c \in C^*$}\label{line:css_coreset_inner_ring}
            \State $\mathbb{C}_c \gets \mathbb{C}_c \backslash R_I(C)$ \Comment{The $\mathbb{C}_c$ denotes the cluster with center point $c$}
            \State $P_0 \gets P_0 \cup  R_I(C)$
            \State $\sigma(c) \gets |R_I(C)|$
            \State $P_1 \gets P_1 \cup \{c\}$
        \EndFor
        
        \For{$c \in C^*$}\label{line:css_coreset_outter_ring}
            \State $\mathbb{C}_c \gets \mathbb{C}_c \backslash ((\mathbb{C}_c \cap (\bigcup_{j \in \Z}G_{j, \min})) \cup (R_O(C) \cap G^O_{\min}))$ 
            \State $P_0 \gets P_0 \cup  (\mathbb{C}_c \cap (\bigcup_{j \in \Z}G_{j, \min})) \cup (R_O(C) \cap G^O_{\min})$
            \State $\sigma(c) \gets |(\mathbb{C}_c \cap (\bigcup_{j \in \Z}G_{j, \min})) \cup (R_O(C) \cap G^O_{\min})|$
            \State $P_1 \gets P_1 \cup \{c\}$
        \EndFor
        
        \State $\beta \gets O(\frac{ \log^2(1/\epsilon)}{2^{O(z \log z)} \cdot \epsilon^2}(k \log |C| + \log\log(1/\epsilon) + \log(1 / \delta)))$
        
        \For{$j \in \Z$ and $j \in (z \log(\epsilon/z), 2 z \log(\epsilon/z))$}\label{line:css_coreset_subcoreset_1}
            \For{$G_{j, b} \not\in G_{j, \min}$}
                \State $D_{j, b} \gets \textsc{GroupSample}(G_{j,b}, \beta, \sigma)$ \Comment{Lemma~\ref{lem:group_sample}, here we store the weight output by the algorithm to $\sigma$ }
            \EndFor
        \EndFor
        
        \State $\beta \gets O(\frac{2^{O(z \log z)} \cdot \log^2(1/\epsilon)}{\epsilon^2}(k \log |C| + \log\log(1/\epsilon) + \log(1 / \delta)))$
        
        \For{$b \in [\max]$}\label{line:css_coreset_subcoreset_2} \Comment{Here the $\max$ denotes the number of interesting outer groups (Definition~\ref{def:group_defs})}
            \State $D_b^O \gets \textsc{SensitivitySample}(G_b^O, \beta, \sigma)$ \Comment{Lemma~\ref{lem:senstivity_sample}, here we store the weight output by the algorithm to $\sigma$}
        \EndFor
        
        \State $D \gets C^* \bigcup_{j,b}D_{j,b} \bigcup_{b}D_b^O$
        \State \Return $(D, \sigma)$
        
    \EndProcedure
    \end{algorithmic}
\end{algorithm}

\subsubsection{Correctness of the Coreset Algorithm}

Note here we do not change the construction of coreset in the algorithm of \cite{css21}. We only plug our center set generation to their algorithm. Thus we have the following correctness theorem and its proof.

\begin{theorem}[Correctness of Algorithm~\ref{alg:coreset_gen_css21}]
\label{thm:correct_css21}
    Let $X \subseteq \R^d$ be a dataset of $n$ points. Let $\epsilon \in (0, 0.1)$ be the coreset parameter. Let $\delta$ be the failure probability. Then the output of Algorithm~\ref{alg:coreset_gen_css21}, a set $D$ together with weight function $\sigma$, is a $(k, \alpha)$-coreset (Definition~\ref{def:eps_coreset}).
\end{theorem}

\begin{proof}
    By Theorem~\ref{thm:center_set_gen_correct} we have that, the set $C^*$ in the algorithm is a $(k, \gamma)$-center set (Definition~\ref{def:alhpa_center_set}). Then by Theorem~1 in \cite{css21}, we have the theorem proved.
\end{proof}

\subsubsection{Running Time Analysis of the Coreset Algorithm}

Here we present the running time of our new coreset generation algorithm, together with the proof of it.

\begin{theorem}[Running time of Algorithm~\ref{alg:coreset_gen_css21}]
\label{thm:runtime_css21}
    Let $X \subseteq \R^d$ be a dataset of $n$ points. Let $\delta$ be the failure probability. Let $\epsilon_0$ be the precision parameter of the \textsc{CenterSetGen} subroutine. 
    Then Algorithm~\ref{alg:coreset_gen_css21} runs in time
    \begin{align*}
        O( \epsilon_0^{-2} z^2 (n k + n d) \log(k/\delta) \log(n/k)).
    \end{align*}
\end{theorem}

\begin{proof}
    The running time of the algorithm can be divided as following parts:
    \begin{itemize}
        \item Line~\ref{line:css_coreset_center_set_gen} takes time  $\mathcal{T}_{\textsc{CenterSetGen}}$ to generate a center set.
        \item Line~\ref{line:css_coreset_sigma_reset} takes time $O(k)$ to reset the weight function $\sigma$.
        \item Line~\ref{line:css_coreset_avrg_cost} takes time $O(n d)$ to compute the average cost of each cluster.
        \item Line~\ref{line:css_coreset_partition} takes time $O(n)$ to partitioning the $X$ into $R_j$.
        \item Line~\ref{line:css_coreset_inner_ring} and Line~\ref{line:css_coreset_outter_ring} takes time $O(k)$ to compute the sets and $\sigma$ respectively.
        \item Line~\ref{line:css_coreset_subcoreset_1} takes time $O(n d)$ to generate the sub-coresets (Note that, here each group $G_{j,b}$ has no intersection with another group, so by Lemma~\ref{lem:group_sample} the time is $O(\sum_{j,b}|G_{j, b}| d) = O(n d)$).
        \item Line~\ref{line:css_coreset_subcoreset_2} takes time $O(n d)$ to generate the sub-coresets (Note that, here each group $G_{j,b}$ has no intersection with another group, so by Lemma~\ref{lem:group_sample} the time is $O(\sum_{b}|G_{b}^O| d) = O(n d)$).
    \end{itemize}
    Adding these above, we have the running time of the algorithm,
    \begin{align*}
        \mathcal{T}_{\textsc{CoresetGen}} 
    =   & ~ \mathcal{T}_{\textsc{CenterSetGen}} + O(k) + O(n d) + O(n) \\
    =   & ~ O( \epsilon_0^{-2} z^2 (n k + n d) \log(k/\delta_0) \log(n/k)).
    \end{align*}
    Thus we complete the proof.
\end{proof}

\section{Data Structure and Algorithm}
\label{sec:data_structure}

In this section, we state our data structure and algorithms formally, together with the analysis of its running time. In Section~\ref{sec:dist_est_ds} we give our Distance Estimation Data Structure and its running time analysis. In Section~\ref{sec:center_set_gen_alg} we state our algorithm for $(k, \alpha)$-center set generation (Definition~\ref{def:alhpa_center_set}) and its running time analysis. In Section~\ref{sec:main_analyze} we state the main algorithm for coreset generation and its running time analysis.

\subsection{Distance Estimation Data Structure}
\label{sec:dist_est_ds}

In this section, we present the distance estimation data structure and its running time analysis. 
\begin{algorithm}[!ht]\caption{Data Structure for Distance Estimation}
\label{alg:data_structure_dist_est}
    \begin{algorithmic}[1]
    \State {\bf data structure} \textsc{DistanceEst} \Comment{Lemma~\ref{lem:ds_dst_est_correct}, Lemma~\ref{lem:ds_dst_est_runtime}}
     \State {\bf members} 
        \State \hspace{4mm} $\Pi \in \R^{m \times d}$ \Comment{The J-L matrix, Lemma~\ref{lem:jl_lemma}}
        \State \hspace{4mm} $\epsilon, \delta \in (0, 0.1)$ \Comment{$\epsilon$ is the precision parameter of estimation, $\delta$ is the failure probability}
        \State \hspace{4mm} $m \in \Z$ \Comment{Size of the sketch}
        \State \hspace{4mm} $n, d \in \Z_+$ \Comment{$n$ is the number of points we maintain, $d$ is the dimension}
        \State \hspace{4mm} $\{v_i\}_{i \in [n]} \subset \R^m$ \Comment{Sketches we maintain}
    \State {\bf end members} 
    \State
    
    \Procedure{Init}{$n \in \Z_+, d \in \Z_+, \delta \in (0, 0.1), \epsilon \in (0,0.1), \{c_i\}_{i \in [n]} \subset \R^d$}
        \State $\epsilon \gets \epsilon, n \gets n, d \gets d, \delta \gets \delta$
        \State $m \gets \Theta(\epsilon^{-2} \log (n/\delta))$
        \State Initialize $\Pi \in \R^{m \times d}$ \label{line:dist_est_init_matrix} \Comment{Lemma~\ref{lem:jl_lemma}}
        \For{$i \in [n]$}
            \State $v_i \gets \Pi c_i$ \label{line:dist_est_init_sketch}
        \EndFor
    \EndProcedure
    \State
    
    \Procedure{Update}{$i \in [n], c \in \R^d$}
        \State $v_i \gets \Pi c$
    \EndProcedure
    \State
    
    \Procedure{Query}{$q \in \R^d$}
        \State $v \gets \Pi q$ \label{line:dist_est_query_sketch}
        \For{$i \in [n]$}
            \State $d_i \gets \|v - v_i\|_2$\label{line:dist_est_query_dist_compute}
        \EndFor
        \State \Return $\{d_i\}_{i \in [n]}$
    \EndProcedure
    \State 
    \Procedure{QueryMin}{$q \in \R^d, z \in \R$}
        \State $v \gets \Pi q$ 
        \For{$i \in [n]$}
            \State $d_i \gets \|v - v_i\|_2^z$
        \EndFor
        \State $j \gets \arg\min_{i\in [n]} d_i$
        \State \Return $j$
    \EndProcedure
    
    \State {\bf end data structure}
    \end{algorithmic}
\end{algorithm}

\begin{lemma}[Correctness for \textsc{Query}]
\label{lem:ds_dst_est_correct}
    Given a set of $n$ points in $\R^d$, an accuracy parameter $\epsilon \in (0, 0.1)$, and a failure probability $\delta \in (0, 0.1)$, the procedure \textsc{Query} of \textsc{DistanceEst} data structure (Algorithm~\ref{alg:data_structure_dist_est}) outputs $\{d_i\}_{i \in [n]}$ satisfing
    \begin{align*}
        \Pr[\forall i \in [n], (1 - \eps) \cdot \|q - c_i\|_2 \le d_i \le (1 + \eps) \cdot \|q - c_i\|_2] \ge 1 - \delta.
    \end{align*}
\end{lemma}

\begin{proof}
    By Lemma~\ref{lem:jl_lemma} we have
    \begin{align*}
        \Pr[(1 - \eps)\|q - c_i\|_2 \le d_i \le (1 + \eps)\|q - c_i\|_2] \ge 1 - \delta/n
    \end{align*}
    for any $i \in [n]$, note that we set $m = O( \epsilon^{-2} \log (n^2/\delta) )$ to get that probability. And by union bound, we have the claim immediately.
\end{proof}

\begin{lemma}[Distance Estimation] 
\label{lem:ds_dst_est_runtime}
    Given a set of $n$ points in $\R^d$, for any accuracy parameter $\epsilon \in (0, 0.1)$ 
    , and any failure probability $\delta \in (0, 0.1)$, there is a data structure (Algorithm~\ref{alg:data_structure_dist_est}) uses $O(\epsilon^{-2} (n + d) \log(n/\delta))$ spaces that supports the following operations
    \begin{itemize}
        \item \textsc{Init}$(n \in \Z_+, d \in \Z_+, \delta \in (0, 0.1), \epsilon \in (0,0.1), \{c_i\}_{i \in [n]} \subset \R^d)$. It takes number of points $n$, dimension $d$, failure probability $\delta$, accuracy parameter $\eps$ and a set of points $\{c_i\}_{i \in [n]}$ as inputs. It runs in time  $O(\epsilon^{-2} n d \log (n / \delta))$,
        \item \textsc{Update}$(i \in [n], c \in \R^d)$. It takes index of point to be update $i$, and an update point $c$ as inputs. 
        It runs in time of $O(\epsilon^{-2} d \log (n / \delta))$,
        \item \textsc{Query}$(q \in \R^d)$. It takes a query point $q$ as input. 
        It runs in time of $O(\epsilon^{-2}(n + d)\log(n / \delta))$.
        \item \textsc{QueryMin}$(q \in \R^d, z \in \R)$. It takes a query point $q$ and a parameter $z$ as power of the distance as input. It runs in time of $O(\epsilon^{-2}(n + d)\log(n / \delta))$.
    \end{itemize}
\end{lemma}

\begin{proof}
    The running time of \textsc{Init} procedure can be divided as follows:
    \begin{itemize}
        \item Line~\ref{line:dist_est_init_matrix} takes time of $O(m d)$ to initialize the matrix.
        \item Line~\ref{line:dist_est_init_sketch} takes time of $O(n m d)$ to compute the sketches.
    \end{itemize}
    Adding them we have the running time of
    \begin{align*}
        O(n m d) = O(\epsilon^{-2} n d \log (n / \delta)).
    \end{align*}
    
    The running time of \textsc{Update} procedure is simply
    \begin{align*}
        O(m d) = O(\epsilon^{-2} d \log (n / \delta)).
    \end{align*}
    
    The running time of \textsc{Query} procedure can be divided as follows.
    \begin{itemize}
        \item Line~\ref{line:dist_est_query_sketch} takes time of $O(m d)$.
        \item Line~\ref{line:dist_est_query_dist_compute} takes time of $O(n m)$.
    \end{itemize}
    Adding them we have the running time of 
    \begin{align*}
        O(m (n + d)) = O(\epsilon^{-2}(n + d)\log(n/\delta)).
    \end{align*}
    Thus we complete the proof.
    
    Then, following the similar reasons of the running time analysis of \textsc{Query}, we get that
    the running time of \textsc{QueryMin} procedure is also $O(\epsilon^{-2}(n + d)\log(n / \delta))$.
    
    For the space, we first store the sketch matrix,
    \begin{align*}
        \text{space for}~\Pi = O(m d) = O(\epsilon^{-2} d \log(n/\delta)).
    \end{align*}
    And we store the sketches,
    \begin{align*}
        \text{space for}~\{v_i\}_{i \in [n]} = O(m n) = O(\epsilon^{-2} n \log(n/\delta)).
    \end{align*}
    Adding them we have the total space of
    \begin{align*}
        O(\epsilon^{-2} (n + d) \log(n/\delta))
    \end{align*}
\end{proof}

\subsection{Center Set Generation Algorithm}
\label{sec:center_set_gen_alg}

In this section, we present our center set generation algorithm and its running time analysis.

\begin{algorithm}[!ht]\caption{Our algorithm for Center Set Generating task}
\label{alg:center_set_gen}
    \begin{algorithmic}[1]
    \Procedure{CenterSetGen}{$\gamma, \delta_0, n, d, U \subset \R^d, k, z$} 
    \Comment{Theorem~\ref{thm:center_set_gen_correct}, Theorem~\ref{thm:center_set_gen_runtime}}
        \State \Comment{$U$ has size of $n$}
        \State $U_0 \gets U$, $S_0 \gets \emptyset$
    	\State Let $\wt{\tau}$ be a map of $U \rightarrow U$
    	\State $\alpha \gets O(1)$, $\beta \gets O(1)$
     	\State \textsc{DistanceEst} $D$
     	\Comment{Algorithm~\ref{alg:data_structure_dist_est}}
     	\State $i \gets 0$
     	\State $V \gets \emptyset$ \Comment{The center set}
    	\While{$|U_i| > \alpha k$} \label{line:center_gen_while_loop}
    	    \State Sample a set $S_i$ from $U_i$ by sampling $\lfloor \alpha k \rfloor$ times, where for each time sample the points with equal probability. \label{line:center_gen_sampling}
    	   \State $D.\textsc{Init}(|S_i|, d, \delta_0, \epsilon_0/z, S_i)$ \label{line:center_gen_dist_init}
    	    \For{$x \in U_i$} 
    	        \State $\wt{d}_x \gets D.\textsc{Query}(x)^z$ \label{line:center_gen_point_query} \Comment{Here we take the $z$-th power for the distance element-wise}
    	    \EndFor
    	    \State Compute $v_i \in \R$ which is the smallest to satisfy $|B(S_i, v_i)| \ge \beta |U_i|$, using the $\wt{d}_x$ computed in last step.\label{line:center_gen_compute_ball} \Comment{Definition~\ref{def:center_set_gen_v_i}}
    	    \State $C_i \gets B(S_i, v_i)$
    	    \For{$x \in C_i$}\label{line:center_gen_loop2}
    	        \State $y \gets $ a point such that $\wt{d}_z(x ,y) \le v_i$ and $y \in S_i$
    	        \State $\wt{\tau}(x) \gets y$
    	    \EndFor
    	    \State $U_{i+1} \gets U_i \backslash C_i$
    	    \State $V \gets V \cup S_i$
    	    \State $i \gets i +1$
    	\EndWhile
    	\For{$x \in U_i$}
    	    \State $\wt{\tau}(x) \gets x$
    	\EndFor
    	\State $V \gets V \cup U_i$
    	\State \Return $\{V, \wt{\tau}\}$
    \EndProcedure
    \end{algorithmic}
\end{algorithm}

Note that in \cite{mp04}, their loop runs for $O(\max\{k, \log n\})$ times. Without loss of generality, we can assume $k = \Omega(\log n)$, so that the loop runs for $O(k)$ times.

\begin{theorem}[Center Set Generation Time]
\label{thm:center_set_gen_runtime}
    Given an $n$-point set in $\R^d$, an accuracy parameter $\epsilon_0 \in (0, 0. 1)$, and a failure probability $\delta_0 \in (0, 0.1)$, the Procedure \textsc{CenterSetGen} (Algorithm~\ref{alg:center_set_gen}) runs in time 
    \begin{align*}
        O( \epsilon_0^{-2} z^2 (n k + n d) \log(k/\delta_0) \log(n/k) ).
    \end{align*}
\end{theorem}

\begin{proof}
    The running time consists of the following parts, and loops them for $O(\log (n/k))$ times:
    \begin{itemize}
        \item Line~\ref{line:center_gen_sampling} takes time of $O(k)$.
        \item Line~\ref{line:center_gen_dist_init} takes time of $O(\epsilon_0^{-2} z^2 |S_i| d \log(k / \delta_0)) = O(\epsilon_0^{-2} z^2 k d \log(k / \delta_0))$.
        \item Line~\ref{line:center_gen_loop2} takes time of $O(n k)$.
        \item Line~\ref{line:center_gen_point_query} takes time of $O(n \epsilon_0^{-2} z^2 (k + d) \log(k/\delta_0))$.
        \item Line~\ref{line:center_gen_compute_ball} takes time of $O(n k)$.
    \end{itemize}
    Adding the above together we get the total running of 
    \begin{align*}
        & ~ \mathcal{T}_{\textsc{CenterSetGen}} \\
    =   & ~ O( \epsilon_0^{-2} z^2 (n k + n d + k d) \log(k/\delta_0) \log(n/k) ) \\
    = & ~ O( \epsilon_0^{-2} z^2 (n k + n d ) \log(k/\delta_0) \log(n/k) )
    \end{align*}
    where the second step follows from $k \leq n$.
    
    Thus we complete the proof.
\end{proof}

\subsection{Our Main Algorithm and its Running Time Analysis}
\label{sec:main_analyze}

In this section, we present our main algorithm(Algorithm~\ref{alg:app:coreset_gen_alg}) and its runtime analysis.

\begin{algorithm}[!ht]\caption{Our algorithm for Coreset generating task}
\label{alg:app:coreset_gen_alg}
    \begin{algorithmic}[1]
    \Procedure{CoresetGen}{$U, n, d, \epsilon, \delta \in (0, 0.1), w, z \ge 1, k \ge 1$} \Comment{Theorem~\ref{thm:main_correct}, and ~\ref{thm:main_runtime}}
        \State $\gamma \gets O(1)$
        \State $C^* \gets \textsc{CenterSetGen}(\gamma,\delta, n, d, U)$\label{line:coreset_center_set_init} \label{line:center_set_gen}\Comment{Algorithm~\ref{alg:center_set_gen}}
        \State \textsc{DistanceEst} $\textsc{de}$
        \State $\textsc{de}.\textsc{Init}(k, d, \delta_1, \epsilon_1/z, C^*)$ \label{line:coreset_dist_init} \label{line:coreset_center_gen} \Comment{Algorithm~\ref{alg:data_structure_dist_est}} 
        \State $N \gets z^{O(z)} \cdot \epsilon^{-\Omega(z)} \cdot k^5 \cdot \log(k/\delta)$
        
        \For{$x \in U$} \label{line:query_loop_sketch_points}
            \State $t \gets \textsc{de}.\textsc{QueryMin}(x,z)$ \label{line:coreset_find_min}
            \State $\wt{c}^*(x) \gets v_t$ \Comment{$\wt{c}^*(x)$ sends $x$ to its approx closest point in $C^*$}
        \EndFor

        \For{$v \in C^*$} 
            \State $X_{v} \gets \{x \in U ~|~  \wt{c}^*(x) = v\}$ \label{line:coreset_loop_set_gen}
        \EndFor
        
        \For{$x \in U$} 
    		    \State $\wt{\sigma}(x) \gets \wt{c}_0 \cdot 2^{2z} \cdot \gamma^2 \cdot (\frac{ \wt{d}_z(x, \wt{c}^*(x))}{ \wt{\cost}_z(X,C^*)} + \frac{1}{|X_{\wt{c}^*(x)}|})$\label{line:coreset_sigma_compute} \Comment{Definition~\ref{def:sigma_func_approx}}
    		\EndFor
    		\State For each $x \in U$, compute $p_x = \frac{\wt{\sigma}(x)}{\sum_{y \in U} \wt{\sigma}(y)}$
    		\State Let $D$ ($|D| = N$) be a subset sampled from $U$, where every $x \in U$ is sampled with prob. $p_x$.  \label{line:coreset_sampling}
    		\For{$x \in D$}
    		    \State $u(x) \gets \frac{\sum_{y \in U} \wt{\sigma} (y)}{|D|\cdot \wt{\sigma}(x)}$ \label{line:coreset_compute_u}
    		\EndFor
    		\State \Return $D$
    \EndProcedure
    \end{algorithmic}
\end{algorithm}

\begin{theorem}[Running time]
\label{thm:main_runtime}
    Let $\epsilon_0, \epsilon_1 \in (0, 0.1)$ be the constant accuracy parameter of the center set generation algorithm and the instance of distance estimation, respectively. Given an $n$-point dataset $U \in \R^d$, a coreset parameter $\epsilon \in (0,0.1)$, and a failure probability $\delta \in (0, 0.1)$, 
    Algorithm~\ref{alg:app:coreset_gen_alg} runs in time
    \begin{align*}
        O( (\epsilon_0^{-2} + \epsilon_1^{-2}) z^2 \cdot (nk + nd) \cdot \log(k /\delta_0) \log(n/k) \log(k /\delta_1) \log(n/\delta_1)).
    \end{align*}
\end{theorem}

\begin{proof}
    The algorithm includes these parts:
    \begin{itemize}
        \item Line~\ref{line:coreset_center_gen} takes $\mathcal{T}_{\textsc{CenterSetGen}}$ to initialize the center set.
        \item Line~\ref{line:coreset_dist_init} takes $O(\epsilon_1^{-2} z^2 k d \log(k/\delta_1))$ to initialize the distance estimation data structure.
        \item Loop the following for $O(n)$ time:
        \begin{itemize}
            \item Line~\ref{line:coreset_find_min} takes $O(\epsilon_1^{-2} z^2 \cdot (k + d) \cdot \log(n/\delta_1))$ to compute the distances.
        \end{itemize}
        \item Line~\ref{line:coreset_loop_set_gen} takes $O(n)$ to generate each set $X_{v_j}$. 
        \item Line~\ref{line:coreset_sigma_compute} takes $O(n)$ to compute the $\wt{\sigma}$. 
    \end{itemize}
    Adding them together we have the total running time of 
    \begin{align*}
            & ~ \mathcal{T}_{\textsc{CoresetGen}}\\
        =   & ~ \mathcal{T}_{\textsc{CenterSetGen}} + O(\epsilon_1^{-2} z^2 k d \log(k/\delta_1)) + O(n \epsilon_1^{-2} z^2 (k + d) \log(n/\delta_1)) \\
        =   & ~ O( (\epsilon_0^{-2} + \epsilon_1^{-2}) z^2 \cdot (nk + nd + kd) \cdot \log(k /\delta_0) \log(n/k) \log(k /\delta_1) \log(n/\delta_1))\\
        =   & ~ O((\epsilon_0^{-2} + \epsilon_1^{-2}) z^2 \cdot (nk + nd) \cdot \log(k /\delta_0) \log(n/k) \log(k /\delta_1) \log(n/\delta_1))
    \end{align*}
    where the first step follows adding them together, the second step follows from Lemma~\ref{thm:center_set_gen_runtime}, the last step follows from $k d \le n d$.
    
    Thus we complete the proof.
\end{proof}




\end{document}